\documentclass[12pt]{iopart}
\usepackage{iopams}
\usepackage[utf8]{inputenc}
\usepackage[english]{babel}
\usepackage{cite}
\expandafter\let\csname equation*\endcsname\relax
\expandafter\let\csname endequation*\endcsname\relax 
\usepackage{amsfonts}
\usepackage{amssymb}
\usepackage{amsmath}
\usepackage{color}
\usepackage{mathrsfs}
\usepackage{relsize}
\usepackage{bm}
\usepackage{hyperref}
\usepackage{soul}
\usepackage{lineno}

\definecolor{dgreen}{rgb}{0.0, 0.2, 0.13}
\definecolor{orange}{rgb}{1.0, 0.27, 0.0}

\newcommand{\beq}{\begin{eqnarray}}
\newcommand{\eeq}{\end{eqnarray}}
\newcommand{\nn}{\nonumber}

\def\keywords#1{\vspace{10pt}
     \begin{indented}
     \item[]\rm Keywords: #1\par
     \end{indented}}
     
\def\AMS#1{\vspace{10pt}
     \begin{indented}
     \item[]\rm AMS classification scheme numbers: #1\par
     \end{indented}}

\usepackage{graphicx}
\usepackage{ragged2e}
\usepackage{relsize}
\usepackage[all]{xy}
\usepackage[mathscr]{euscript}
\let\oldnormalfont\normalfont
\def\normalfont{\oldnormalfont\mdseries}

\providecommand{\fibbun}[2]{\pi_{#2\! #1}}

\begin{document}

\title{Covariant momentum map for non-Abelian topological BF field theory}
\author{Alberto Molgado$^{1,2}$ and Ángel Rodríguez--López$^{1}$}

\address{$^{1}$ Facultad de Ciencias, Universidad Autonoma de San Luis 
Potosi \\
Campus Pedregal, 
Av.~Parque Chapultepec 1610,
Col.~Privadas del Pedregal,
San Luis Potosi, SLP, 78217, Mexico}
\address{$^{2}$ Dual CP Institute of High Energy Physics, Colima, Col, 28045, Mexico}

\eads{\mailto{\textcolor{blue}{alberto.molgado@uaslp.mx}},\ \mailto{\textcolor{blue}{angelrodriguez@fc.uaslp.mx}}}

\begin{abstract}
We analyze the inherent symmetries associated to the non-Abelian
topological BF theory from the geometric and covariant perspectives of
the Lagrangian and the multisymplectic formalisms. 
At the Lagrangian level, we classify the symmetries of the theory as natural and Noether symmetries and construct the associated Noether currents, while at the multisymplectic level the symmetries of the theory arise as covariant canonical transformations.   These transformations 
allowed us to build  within the multisymplectic approach,  in a complete covariant way, the momentum maps 
which are analogous to the conserved Noether currents.  The covariant 
momentum maps are fundamental to recover, after the space plus time 
decomposition of the background manifold, not only the extended 
Hamiltonian of the BF theory but also the generators of the gauge 
transformations which arise in the instantaneous Dirac-Hamiltonian analysis 
of the first-class constraint structure that characterizes the BF model under 
study.  
To the best of our knowledge, this is the first non-trivial physical model associated to General Relativity for which both natural and Noether symmetries have been analyzed at the multisymplectic level. Our study 
shed some light on the understanding of the manner in which the  generators of gauge transformations may be recovered from 
the multisymplectic formalism for field theory. 
\end{abstract}



\AMS{70S05, 70S15, 37J15, 70G65, 70G65}

\section{Introduction}

BF theories are a class of diffeomorphism invariant non-metric topological field theories 
characterized by the absence of local degrees of freedom~\cite{Horowitz, Cattaneo, Birmingham}. As it is well known, these kind of field theories have a strong relationship with Einstein theory of General Relativity since, following the work developed by Plebanski \cite{Plebanski}, it is possible to write General Relativity as a constrained BF theory.  This relation gives rise to the so-called BF gravity 
models~\cite{Speziale, Montesinos2}. Recently, BF theories have been exhaustively explored in the literature from different perspectives and approaches at both the classical and the quantum levels 
\cite{Horowitz, Cattaneo, Sardanashvily1, Birmingham, Spinfoams1, Spinfoams2, DeGracia, Montesinos3, Angel}, mainly due to the features that distinguish these topological field theories which provide an extended framework 
to explore different aspects of gauge theories. In this context, of particular interest to us is the case of the $4$-dimensional non-Abelian topological BF theory, which at the classical level has been studied from different perspectives including the Lagrangian \cite{Cattaneo, Montesinos1}, the pure Dirac-Hamiltonia~\cite{Escalante1}, the Hamilton-Jacobi~\cite{DeGracia}, and the covariant canonical~\cite{Montesinos3} formalisms.

From our own perspective, a particular 
relevant way to analyze BF theories is by 
considering the instantaneous Dirac-Hamiltonian 
formulation as it reveals the inherent gauge symmetry 
of the theory, as described in~\cite{Montesinos1, Escalante1}.
This instantaneous Dirac-Hamiltonian formulation for classical field theory initiates by
introducing a foliation of the space-time into space-like hypersurfaces, and by considering the dynamical variables of the theory as functions on those hypersurfaces that are characterized as the space of Cauchy data.  
In this manner, time evolution off the hypersurface is obtained through Hamilton's 
equations.  However, 
as it is well known, this formulation 
conceals the explicit covariant nature of a field theory.  As a consequence, 
at the quantum level, by imposing 
an appropriate canonical quantization, the instantaneous Dirac-Hamiltonian formulation 
gives rise to non-covariant models of quantum field theory~\cite{Forger1}. 
Bearing this in mind, we consider the analysis of the the gauge symmetries for BF theories 
within the multisymplectic formulation for field theories.  The multisymplectic approach provides a finite-dimensional and geometric covariant Hamiltonian-like formulation for classical field theory \cite{GIMMSY1, Crampin, Gotay1, DeLeon1}, 
which is based on the De Donder–Weyl canonical theory.  The multisymplectic formalism starts by introducing the definition of the multimomenta phase-space, that is, a finite-dimensional manifold locally constructed by associating to each field variable of the theory a set of  multimomenta variables that correspond to a covariant extension of the standard Dirac-Hamiltonian momenta defined by considering temporal derivatives of the fields~\cite{Forger1}. This multimomenta phase-space is endowed with a multisymplectic form which plays a very relevant role within this covariant formulation, since it allows 
not only to obtain the classical field equations but also to 
to describe the symmetries of the classical field theory. In particular, within the multisymplectic approach the symmetries of a classical field theory are understood as covariant canonical transformations, that is, transformations on the multimomenta phase-space that leave invariant the multisymplectic form, thus allowing us to extend, for the case of  infinitesimal symmetries, the first Noether's theorem to the multisymplectic framework and,
in consequence, we are able to obtain 
the so-called covariant momentum maps, which are the analogous on the multimomenta phase-space of the Noether currents on the ordinary 
instantaneous phase-space \cite{GIMMSY1, Forger2, DeLeon1}.
In addition, under certain technical conditions we may choose a Cauchy surface in such a manner that the covariant momentum maps of a given classical field theory project  to functions into the instantaneous phase-space (Cauchy data space) of the theory, which coincide with the instantaneous Hamiltonian and the genuine momentum maps of the theory in the standard canonical Dirac-Hamiltoninan formulation~\cite{GIMMSY1, GIMMSY2, Fischer}. In particular, for the case of singular Lagrangian systems~\cite{QGS}, the projected momentum map associated to the gauge symmetry group of a classical field theory coincides with the generator of the infinitesimal gauge transformations of the system in the instantaneous Dirac-Hamiltonian analysis, at least for the physical models explored 
in the literature~\cite{GIMMSY1, GIMMSY2, DeLeon1}. Furthermore, by the second Noether's theorem (see, for instance, \cite{Lee, Avery}), for any solution of the field equations of the theory, the covariant momentum map integrated over a Cauchy surface is zero, thus establishing that the projected momentum map associated to the gauge symmetry group of the theory must vanish on the set of admissible Cauchy data for the evolution equations.  Consequently,  the zero level set of the projected momentum map of the gauge symmetry group theory must coincide with the first class constraints of the system in the Dirac-Hamiltonian approach, as described 
in~\cite{GIMMSY1, GIMMSY2, Fischer}. In that 
sense, starting from the multisymplectic framework, we have a natural way to recover within the instantaneous Dirac-Hamilton approach the constrained and gauge structure of a given  
singular Lagrangian field theory.
 
Our main propose is thus to analyse the $4$-dimensional non-Abelian topological BF theory within the multisymplectic framework. This in order to describe the features of this diffeomorphism invariant topological field theory within a finite-dimensional, geometric and covariant framework. Our analysis starts by describing the non-Abelian BF theory within a geometric Lagrangian approach, where the field variables of the model are understood as sections of a vector fibre-bundle of differential forms on the space-time manifold on which the field theory is defined. Within this Lagrangian approach, we classify the symmetries of the theory in natural and Noether symmetries and obtain the associated field equations of the system, which allow us to recover the Noether currents of the theory as reported in the literature~\cite{Montesinos1}. Then, we proceed to construct the multimomenta phase-space and introduce the associated multisymplectic form of the BF theory for which the symmetries of the model appear as covariant canonical transformations.  Besides, we also construct the covariant momentum maps associated to the BF model. Finally, following the works developed in \cite{GIMMSY1, GIMMSY2, Fischer, DeLeon1, DeLeon2}, after performing the space plus time decomposition for the BF theory, we are able to recover not only the extended Hamiltonian, but also the generator of the infinitesimal gauge transformations as well as the first and second class constraints of the non-Abelian topological BF theory, which explicitly correspond to the results obtained by means of the Dirac's algorithm in references~\cite{Escalante1, DeGracia}. 

The rest of the paper is organized as follows: in Section \ref{Multisec} we briefly describe the Lagrangian and the multisymplectic covariant formalisms for a generic classical field theory, paying  special attention to the study of symmetries associated to both covariant approaches. We also introduce a brief description of the space plus time decomposition for a classical field theory that will allow us to connect the multisymplectic and the instantaneous Dirac-Hamiltonian formulations. In Section \ref{sec:BFTheory} we present a description of the $4$-dimensional non-Abelian topological BF field theory introducing both
covariant formulations, namely the Lagrangian and multisymplectic.  For both  approaches we 
discuss in detail the associated field equations, classify the symmetries and construct the associated Noether currents and covariant momentum maps of the BF model of our interest.  In this section, we also perform the space plus time decomposition of the BF theory at both, the Lagrangian and the multisymplectic levels, which will allow us to recover the Dirac-Hamiltonian analysis of the non-Abelian topological BF field theory. Finally, in Section~\ref{sec:Conclu} we introduce some concluding remarks.

\section{Covariant formalisms for classical field theory}\label{Multisec}

In this section we briefly describe the Lagrangian and the multisymplectic formalisms, which are finite-dimensional and covariant approaches for classical field theory. In particular, we will focus our attention on the study of the symmetries of a classical field theory within these frameworks, which will allow us to construct the Noether currents, in the Lagrangian case, and the covariant momentum maps, within the multisymplectic approach, associated to the theory.  Our main motivation to do this is to recover the constraints and the extended Hamiltonian 
of the field theory taking as a starting point the multisymplectic formulation. In consequence, we give a brief description of the process to perform the space plus time decomposition for classical field theory at both the Lagrangian and the multisymplectic levels.  This will allow us to recover, on the space of Cauchy data, the instantaneous Dirac-Hamiltonian analysis of the theory where, as we will see below, the first class constraints of the system, within Dirac's terminology, will emerge by means 
of the vanishing of the projected covariant momentum maps of the theory on the space of Cauchy data.  In what follows, we will review the main ideas 
of the multisymplectic formalism and the associated momentum maps as 
developed in references~\cite{Sardanashvily1, GIMMSY1, GIMMSY2, Forger1,
Crampin, DeLeon1, DeLeon2}.  We would like to encourage the reader to examine those 
references for further details and technicalities. 

\subsection{Covariant configuration space}

Before proceeding with the geometric descriptions for classical field theory, we will introduce a set of notations that we will use along this paper. To start, let us consider $E$ a finite-dimensional smooth manifold, such that, given an arbitrary point $e\in E$, we denote by $T_{e}E$ and $T^{*}_{e}E$ the tangent and cotangent spaces of $E$, respectively. We also introduce ${\Lambda^{q}}_{e}E$ to be the vector space of differential $q$-forms at the point $e\in E$, while the space of smooth vector fields on $E$ will be denoted by $\mathfrak{X}\left(E\right)$.  We will consider these basic structures for any of the manifolds to be considered below.

Now, let $\left(Y, \fibbun{Y}{X},  X\right)$ be a finite-dimensional fibre-bundle, where $X$ stands for the base space, which is an $n$-dimensional smooth manifold without boundary and local coordinates denoted by $\{x^{\mu}\}_{\mu=0}^{n-1}$, while $\fibbun{Y}{X}:Y\rightarrow X$ is the standard projector map, and $Y$ is the so-called total space whose fibres are $m$-dimensional smooth manifolds locally represented by $\{y^{a}\}_{a=1}^{m}$. In what follows, we will refer to the fibre-bundle $\left(Y, \fibbun{Y}{X}, X\right)$ just as $\fibbun{Y}{X}$, details about the fibre-bundle formalism may be found in \cite{Saunders, Sardanashvily1}. For an arbitrary point $x \in X$, we introduce $Y_{x}$ to be the fibre $\fibbun{Y}{X}^{-1}(x)$ of $Y$ and denote by $\mathscr{Y}$ the space of all sections of $\fibbun{Y}{X}$, such that, given the fibre coordinates $\left(x^{\mu}, y^{a}\right)$ at the point $y\in Y$, a local section $\phi\in\mathscr{Y}$ can be represented on $Y$ by means of the composition function $y\circ \phi=\left(x^{\mu},\phi^{a}\right)$. Thus, given a classical field theory, the fibre-bundle $\fibbun{Y}{X}$ will be the covariant configuration space of the theory where the physical fields will be sections of this bundle \cite{GIMMSY1}.

In this formulation of classical field theory, the analogous to the tangent bundle of classical mechanics is the first order jet bundle, $\fibbun{J^{1}\!Y}{Y}:J^{1}Y \rightarrow Y$, associated to $\fibbun{Y}{X}$. In order to define it, let us consider the points $y\in Y$ and $x=\fibbun{Y}{X}(y)\in X$, such that, the fibre $J^{1}_{y} Y$ of $\fibbun{J^{1}\!Y}{Y}$ at $y\in Y$ is defined by $J^{1}_{y}Y:=\{\gamma\in L\left(T_{x} X,T_{y}Y\right)\left| \right.\, ({\fibbun{Y}{X}})_{\ast}\circ\gamma=\mathrm{Id}_{T_{x}X}\}$, that is, the set of all linear mappings $\gamma:T_{x} X\rightarrow T_{y}Y $, where $({\fibbun{Y}{X}})_{\ast}:TY\rightarrow T X$ stands for the differential map associated to $\fibbun{Y}{X}$. Since $J^{1}_{y}Y$ is an affine space modelled on the vector space $L\left(T_{x} X, V_{y}Y\right)$ with $V_{y}Y$ denoting the vertical tangent bundle of $Y$ at $y\in Y$, the fibre-bundle $\fibbun{J^{1}\!Y}{Y}$ can be regard as an affine bundle locally represented by $\gamma:=(x^{\mu},y^{a}, y^{a}_{\mu})$ \cite{Saunders, Forger1}. In particular, we have that, given a section $\phi\in\mathscr{Y}$ at $x\in X$, the map $\phi_{\ast}:T_{x} X\rightarrow T_{\phi(x)}Y$ is an element of $J^{1}_{\phi(x)}Y$, which allows to identify $j^{1}\phi:=\phi_{\ast}$, the 1-jet prolongation of $\phi$, as a section of the fibre-bundle $\fibbun{J^{1}\!Y}{X}:J^{1}Y\rightarrow X$ at $x\in X$, whose local representation on $J^{1}Y$ is explicitly given by the composition $\gamma\circ j^{1}\phi=\left(x^{\mu}, \phi^{a}, \phi^{a}_{\mu}\right)$. Here we have adopted the notation $\phi^{a}_{\mu}:=\partial_{\mu}\phi^{a}$. As we will see bellow, the fibre-bundle formalism provides a natural framework to study classical field theory, as discussed in \cite{Saunders, Sardanashvily1, Forger1, Crampin, DeLeon1}.

\subsection{Symmetries in the Lagrangian formulation}\label{Lagsec}

Now, we will introduce the basic objects to study classical field theory within the Lagrangian formalism in the jet bundle approach. In particular, we will focus our attention on describing symmetries of classical field theory in this geometric formulation, which will be relevant to us in order to extend Noether's theorems to the multisymplectic framework, as we will see in subsection~\ref{SCMPS}.  

To begin, let $\left(Y,\fibbun{Y}{X},X\right)$ be the covariant configuration space of a classical field theory defined by the action principle
\begin{equation}\label{APFT}
\mathcal{S}[\phi]:=\int_{ X} \left( j^{1}\phi\right)^{*}\mathcal{L}\left(\gamma\right)\, ,
\end{equation} 
where $\mathcal{L}:J^{1}Y\rightarrow \Lambda^{n} X$ stands for the Lagrangian density of the theory, which can be locally given by  $\mathcal{L}:=L(\gamma)\,d^{n}x$, with $L:J^{1}Y\rightarrow\mathbb{R}$ denoting the Lagrangian function of the system. In order to describe the dynamics of the generic field theory \eqref{APFT}, 
we will start by introducing the so-called Poincar\'e--Cartan $n$-form, $\Theta^{(\mathcal{L})}\in\Lambda^{n} J^1Y$, that is locally defined as
\begin{equation}\label{PCF}
\Theta^{(\mathcal{L})}:= \frac{\partial L}{\partial y^{a}_{\mu}}dy^{a}\wedge d^{n-1}x_{\mu}+\left(L-\frac{\partial L}{\partial y^{a}_{\mu}}y^{a}_{\mu}\right)\, d^{n}x\, .
\end{equation}
In a similar manner, we will introduce the $(n+1)$-form
$\Omega^{(\mathcal{L})}:= 
-d\Theta^{(\mathcal{L})}\in\Lambda^{n+1} J^1Y$ which explicitly reads
\beq
\Omega^{(\mathcal{L})}=
dy^{a}\wedge d\left(\frac{\partial L}{\partial y^{a}_{\mu}}\right)\wedge d^{n-1}x_{\mu}-d\left(L-\frac{\partial L}{\partial y^{a}_{\mu}}y^{a}_{\mu}\right)\wedge d^{n}x\, .
\eeq 
This $(n+1)$-form is necessary in order to obtain the correct 
field equations at the Lagrangian level.  Indeed, considering a local 
section $\phi\in\mathscr{Y}$ as a critical point of the action principle \eqref{APFT} and $V\in\mathfrak{X}(J^{1}Y)$ an arbitrary projectable vector field on $J^{1}Y$, that is, a vector field that projects by means of the differential map ${\fibbun{J^{1}\!Y}{Y}}_{\ast}$ into a well defined vector field on $Y$, it is possible to write the following condition
\begin{equation}\label{ELFE}
\left(j^{1}\phi\right)^{*}\left(V\lrcorner\, \Omega^{(\mathcal{L})} \right)=0\, ,
\end{equation}
which may be shown to be completely equivalent to the Euler--Lagrange field equations of the theory~\cite{Sardanashvily1, GIMMSY1, DeLeon1, Fischer}, namely
\begin{equation}
\frac{\partial L}{\partial y^{a}} \circ\!\left(j^{1}\phi\right)-\frac{\partial}{\partial x^{\mu}}\left(\frac{\partial L}{\partial y ^{a}_{\mu}}\circ\!\left(j^{1}\phi\right)\right)=0\,.
\end{equation}

As previously mentioned, we want to study symmetries of classical field theory within this covariant geometric framework, and in particular, the action of a Lie group on the covariant configuration bundle of the theory. To this end, let us consider a Lie group $\mathcal{G}$ whose Lie algebra is denoted by $\mathfrak{g}$, which acts on $ X$ and $Y$ by diffeomorphisms and $\fibbun{Y}{X}$-bundle automorphisms, respectively (see~\cite{Sardanashvily1} for further details on fibre-bundle maps). Now, let $\eta\in\mathcal{G}$, we denote by $\eta_{X}$ and $\eta_{Y}$ the corresponding transformations of $X$ and $Y$ associated to the group element $\eta$, respectively. Besides, we denote by $\xi_{\eta}\in\mathfrak{g}$ the infinitesimal generator of $\eta$, such that $\xi_{\eta}^X:=\xi^{\mu}(x)\partial_{\mu}\in\mathfrak{X}\left( X\right)$ and $\xi_{\eta}^{Y}:=\xi^{\mu}(x)\partial_{\mu}+\xi^{a}(x,y)\partial_{a}\in\mathfrak{X}\left(Y\right)$ stand for the corresponding infinitesimal generators of the transformations $\eta_{X}$ and $\eta_{Y}$, respectively. Naturally, $\mathcal{G}$ acts on $J^{1}Y$ by $\eta_{J^{1}\!Y}(\gamma):T_{\eta_{X}(x)} X\rightarrow T_{\eta_{Y}(y)}Y$ for $\gamma\in J ^{1}Y$, which is defined as $\eta_{J^{1}\!Y}(\gamma):={\eta_{Y}}_{*}\circ \gamma\circ {\eta_{X}}_{*}^{-1}$ and locally reads
\begin{equation}
\eta_{J^{1}\!Y}(\gamma)=\left(\eta^{\mu}_{X}(x), \eta^{a}_{Y}(x,y), \left(\partial_{\nu}\eta ^{a}_{Y}(x,y)+\partial_{b}\eta^{a}_{Y}(x,v)y^{b}_{\nu}\right)\partial_{\mu}\left(\eta^{-1}_{X}(x)\right)^{\nu}\right)\, ,
\end{equation}
where $\partial_{b}:=\partial/\partial y^{b}$ denotes the partial derivative with respect to the coordinates $y^{a}$, while the jet prolongation $\xi_{\eta}^{J^{1}Y}:=j^{1}\xi_{\eta}^{Y}\in\mathfrak{X}\left(J^{1}Y\right)$  associated to vector field $\xi_{\eta}^{Y}$ is locally represented by 
\begin{equation}\label{PJB}
\xi_{\eta}^{J^{1}\!Y}=\xi^{\mu}(x)\partial_{\mu}+\xi^{a}(x,y)\partial_{a}+\left(\partial_{\mu}\xi^{a}(x,y)+\partial_{b}\xi^{a}(x,y)y^{b}_{\mu}-\partial_{\mu}\xi^{\nu}(x)y^{a}_{\nu}\right)\partial^{\mu}_{a}\, ,
\end{equation}
which corresponds to the vector flow associated to the infinitesimal transformation $\eta_{J^{1}\!Y}$ on $J^{1}Y$. Here, we have introduced the notation $\partial^{\mu}_{a}:=\partial / \partial y^{a}_{\mu}$ \cite{ Saunders, Sardanashvily1, GIMMSY1}.

With all this in mind, we are in the position to relate the above description of fibre-bundle morphisms with the symmetries of a classical field theory. To start, let $\mathcal{G}$ be a Lie group acting on $Y$ by $\fibbun{Y}{X}$-bundle automorphisms.  If this group leaves the 
physical action\eqref{APFT} invariant, we say that $\mathcal{G}$ is the symmetry group associated to the theory \cite{GIMMSY1, DeLeon1}. In particular, we are interested in studying two types of symmetries, the so-called natural and divergence symmetries.  On the one hand, we say that the infinitesimal transformation $\eta_{J^{1}\!Y}$ corresponds to a natural symmetry of the theory, if the following condition holds
\begin{equation}
\label{eq:natural}
\mathfrak{L}_{\xi_{\eta}^{J^{1}Y}}\Theta^{(\mathcal{L})}=0\, ,
\end{equation}
where $\mathfrak{L}$ denotes the Lie derivative on $J^{1}{Y}$ along the vector flow $\xi_{\eta}^{J^{1}\!Y}\in\mathfrak{X}\left(J^{1}Y\right)$ of $\eta_{J^{1}\!Y}$ applied to the Poincaré--Cartan form $\Theta^{(\mathcal{L})}\in \Lambda^{n}J^{1}Y$. On the other hand, we say that $\eta_{J^{1}\!Y}$ is a divergence or Noether symmetry of the theory  if $\xi_{\eta}^{J^{1}\!Y}$ satisfies
\begin{equation}
\label{eq:noethersym}
\mathfrak{L}_{\xi_{\eta}^{J^{1}\!Y}}\Theta^{(\mathcal{L})}=d\alpha\, ,
\end{equation}
being $\alpha\in\Lambda^{n-1} J^{1}Y$ a semibasic differential $(n-1)$-form on $J^{1}Y$,
that is, for $\gamma\in J^1 Y$ and $y=\fibbun{J^1 Y}{Y}(\gamma)\in Y\subset J^1 Y$, $\alpha$ is an element of the subset 
$ \Lambda^{n-1}_{2}Y:=\{a\in \Lambda^{n-1}_{y}Y\left|\right.\, V\lrcorner\, W\lrcorner \,a =0,~\forall~~ V,W\in V_{y}Y\, \}\subset \Lambda^{n-1}_\gamma J^{1} Y$,
where $V_{y}Y$ stands for the vertical tangent bundle over $Y$~\cite{Crampin, DeLeon1}. Bearing in mind the symmetry definitions above and taking into account condition~\eqref{ELFE}, we have that for a solution of the 
Euler-Lagrange field equations $\phi$ and an infinitesimal natural or divergence symmetry of the theory $\xi_{\eta}^{J^{1}\!Y}$, 
the quantity given by 
\begin{equation}
\mathcal{J}^{(\mathcal{L})}\!\left(\xi_{\eta}\right)=(j^{1}\phi)^{*}\left( \xi_{\eta}^{
J^{1}Y}\lrcorner\,\Theta^{(\mathcal{L})}-\alpha\right)\, ,
\end{equation} 
corresponds to a Noether current and therefore to a conserved quantity of the theory~\cite{Sardanashvily1, GIMMSY1, Crampin, DeLeon1, Fischer}. As we will see below, this geometric description of the Lagrangian formulation for classical field theory will allow us to extend the first Noether's theorem to the multisymplectic formalism. This is fundamental in order to construct conserved quantities within this last framework. Following \cite{GIMMSY1, GIMMSY2, Gotay1}, we will denote this extension as the covariant momentum maps and it will play a fundamental role in our analysis
of the symmetries of a classical field theory in the subsequent sections.

\subsection{Covariant multimomenta phase-space}
\label{sec:multi}

In order to introduce a covariant Hamiltonian-like formulation for classical field theory, we need to define the dual jet bundle $\fibbun{J^{1}\!Y^{\star}}{Y}:J^{1}Y^{\star}\rightarrow Y$, that is, the dual bundle associated to the jet bundle $\fibbun{J^{1}\!Y}{Y}$ ( see \cite{Saunders, Sardanashvily1} for details about dual bundles). In order to define it, let us consider the points $y\in Y$ and $x=\fibbun{Y}{X}(y)\in X$, such that, the fibre $J^{1}_{y}Y^{\star}$ at $y\in Y$ is defined to be the vector space $J^{1}_{y}Y^{\star}:=\mathrm{Aff}\left(J_{y}^{1}Y, \Lambda^{n}_{x} X\right)$, that is, the set of all affine maps $z^{\star}:J_{y}^{1}Y\rightarrow\Lambda^{n}_{x} X$, where the dual pairing of $J^{1}Y^{\star}$ with $J^{1}Y$ is given by
\begin{equation}
\begin{aligned}
\left\langle~\cdot\,,\,\cdot~\right\rangle~ :~&J^{1}Y^{\star}\times_{Y} J^{1}Y \rightarrow \Lambda^{n} X\, ,
\end{aligned}
\end{equation}
with $\times_{Y}$ denoting the fibre-bundle product over $Y$, as described in  \cite{Sardanashvily1, Forger1}. In particular, the dual jet bundle $J^{1}Y^{\star}$ can be locally represented by the coordinates $z^{\star}:=\left(x^{\mu},y^{a}, p, p^{\mu}_{a}\right)$, which are defined by requiring the dual pairing between the points $z^{\star}\in J^{1}Y^{\star}$ and $\gamma\in J^{1}Y$ to be given by $\left\langle z^{\star},\gamma \right\rangle=\left(p^{\mu}_{a}y^{a}_{\mu}+p\right)\!d^{n}x$, as discussed in \cite{GIMMSY1, Forger1, DeLeon2}. Now, let us consider the vector fibre-bundle $\fibbun{Z}{Y}:Z\rightarrow Y$, where, in analogy to the Lagrangean case, $Z:=\Lambda^{n}_{2}Y$ is the space of semibasic differential $n$-forms on $Y$, that is, the subbundle $Z\subset \Lambda^{n}Y$ whose fibre $ Z_{y}$ at $y\in Y$ is defined by $Z_{y}:=\{\text{\Large{$z$}}\in \Lambda^{n}_{y}Y\left|\right.\, V\lrcorner\, W\lrcorner\, \text{\Large{$z$}} =0,~\forall~~ V,W\in V_{y}Y\, \}$.  
From the definition, the elements of $ Z$ can be locally written as
\begin{equation}
\text{\Large{$z$}}:= p\, d^{n}x+p^{\mu}_{a}dy^{a}\wedge d^{n-1}x_{\mu}\, ,
\end{equation}
where we have defined $d^{n-1}x_{\mu}:=\partial_{\mu}\lrcorner\,d^{n}x$, such that, $\fibbun{Z}{Y}$ can be locally represented by $z:=\left(x^{\mu},y^{a}, p, p^{\mu}_{a}\right)$, giving rise to the fibre-bundle isomorphism $\Phi: Z\rightarrow J^{1}Y^{\star}$, which can be defined by $\left\langle \Phi(z),\gamma\right\rangle=\gamma^{*}z\in \Lambda^{n}_{x} X$, where $z\in Z_{y}$, $\gamma\in J^{1}_{y}Y$ and $x=\fibbun{Y}{X}(y)\in X$ for any $y\in Y$, respectively.

The main idea of introducing the vector space $Z\subset \Lambda^{n} Y$ stands for the possibility to construct canonical forms on it, analogous to the canonical $1$ and $2$-forms in classical mechanics, and 
then we may use the isomorphism $\Phi$ to induce those canonical forms in the dual jet bundle $J^{1}Y^{\star}$. In fact, the  vector space $ Z$ is endowed with a canonical $n$-form $\Theta\in\Lambda^{n} Z$, that we will 
refer to as the multisymplectic potential, defined by 
\begin{equation}
\label{eq:potential}
\Theta:=p\, d^{n}x + p^{\mu}_{a}dy^{a}\wedge d^{n-1}x_{\mu}\, ,
\end{equation}
which induces on $ Z$ the so-called multisymplectic $(n+1)$-form, $\Omega:=-d\Theta\in\Lambda^{n+1}Z$, given by 
\begin{equation}
\label{eq:omega}
\Omega=dy^{a}\wedge dp^{\mu}_{a}\wedge d^{n-1}x_{\mu}-dp\wedge d^{n}x\, .
\end{equation} 
In consequence, the pair $\left( Z,\Omega\right)$ is denominated the covariant multimomenta phase-space \cite{GIMMSY1, Forger1, Crampin, DeLeon1}, that will play a relevant part of our study in what follows.

Additionally, we may introduce the covariant Legendre transformation,
$\mathbb{F}\mathcal{L}:J^{1}Y\rightarrow J^{1}Y^{\star}\cong Z$, locally 
defined by
\begin{equation}\label{CLT}
\mathbb{F}\mathcal{L}(\gamma)=\left(x^{\mu}, y^{a}, L-\frac{\partial L}
{\partial y^{a}_{\mu}}y^{a}_{\mu}, \frac{\partial L}{\partial y^{a}
_{\mu}}\right)\, ,
\end{equation} 
which, in analogy to classical mechanics, relates the Lagrangian and multisymplectic formalisms.  From our point of view, the covariant Legendre transformation is 
relevant as it allows us to obtain by pull-back the Poincaré--Cartan $n$-form, $\Theta^{(\mathcal{L})}=\mathbb{F}\mathcal{L}^{*}\Theta\in\Lambda^{n}J^{1}Y$, and  the $(n+1)$-form $\Omega^{(\mathcal{L})}=\mathbb{F}\mathcal{L}^{*}\Omega\in\Lambda^{n+1}J^{1}Y$, respectively.  We will use 
this covariant Legendre transformation in order to induce the information 
on the symmetries of a system in either formalism.

\subsection{Symmetries in the covariant multimomenta phase-space}
\label{SCMPS}

In this subsection, we will discuss the symmetries of a classical field theory within the covariant multisymplectic approach. Strictly speaking, we will study the action of the symmetry group $\mathcal{G}$ of a given classical field theory on its associated covariant multimomenta phase-space $\left( Z,\Omega\right)$. In particular, we want to classify the action of $\mathcal{G}$ on $\left( Z,\Omega\right)$ in a similar fashion to the Lagrangian case. To this end, we will introduce the notion of a covariant canonical transformation that will allow us to construct, in the case of infinitesimal symmetries, the covariant momentum maps on $Z$, which are the analogous to the Noether currents within the  covariant 
multisymplectic approach. 

To start, let us consider the covariant phase-space $\left( Z,\Omega\right)$ associated to the covariant configuration space $\left(Y,\fibbun{Y}{X},X\right)$ of a classical field theory. Let $\tilde{\eta}_{Z}: Z\rightarrow Z$ be a bundle automorphism with an associated diffeomorphism $\tilde{\eta}_{X}$ on $ X$. Thus we will say that $\tilde{\eta}_{Z}$ is a covariant canonical transformation if it satisfies the identity
\begin{equation}
{\tilde{\eta}_{Z}}^{*}\Omega=\Omega,
\end{equation}
where $\Omega$ is the multisymplectic $(n+1)$-form on $Z$ defined 
in~\eqref{eq:omega}. Additionally, we say that $\tilde{\eta}_{Z}$ is a special covariant canonical transformation if the following condition holds
\begin{equation}
\label{eq:SpecialCov}
{\tilde{\eta}_{Z}}^{*}\Theta=\Theta,
\end{equation}
where $\Theta$ is the multisymplectic potential~\eqref{eq:potential}, as 
described in~\cite{GIMMSY1}.

Next, in order to establish an analogy with symmetries in the Lagrangean case we require to introduce infinitesimal covariant canonical transformations 
within the multisymplectic approach.  To start, let $\chi_{Y}$ be a $\fibbun{Y}{X}$-automorphism
with an associated diffeomorphism $\chi_{X}$ on $ X$. The canonical lift of $\chi_{Y}$ to $ Z$, $\chi_{Z}: Z\rightarrow Z$, is defined as $\chi_{Z}(z):=\left({\chi_{Y}}^{-1}\right)^{*}z$ for $z\in Z$, which implies that canonical lifts are special covariant canonical transformations \cite{GIMMSY1}. In particular, if $\chi_{Y}$ is the vector flow associated to the vector field ${\tilde{\xi}_{\chi}}^{Y}
\in\mathfrak{X}\left(Y\right)$, then the vector field on $ Z$ defined by
\beq
{\tilde{\xi}_{\chi}}^{Z}
& := & 
\tilde{\xi}^{\mu}(x)\partial_{\mu}+\tilde{\xi}^{a}(x,y)\partial_{a}-\left(p\,\partial_{\nu}
\tilde{\xi}^{\nu}(x)+p^{\nu}_{b}\partial_{\nu}\tilde{\xi}^{b}(x,y)\right)\partial_{p}
\nn\\
& & 
+\left(p^{\nu}_{a}\partial_{\nu}\,\tilde{\xi}^{\mu}(x)-p^{\mu}_{b}\partial_{a}\tilde{\xi}^{b}(x,y)-p^{\mu}_{a}\partial_{\nu}\tilde{\xi}^{\nu}(x)\right)\partial^{a}_{\mu}\, ,
\label{CLMMPS}
\eeq
corresponds to the infinitesimal generator of $\chi_{Z}$, that is, the canonical lift of ${\tilde{\xi}_{\chi}}^{Y}$ to $Z$, where we have introduced the short notations $\partial_{p}:=\partial/\partial p$ and $\partial^{a}_{\mu}:=\partial/\partial p^{\mu}_{a}$. Thus, from the definition and by direct calculation, ${\tilde{\xi}_{\chi}}^{Z}$ satisfies the condition $\mathfrak{L}_{{\tilde{\xi}_{\chi}}^{Z}}\Theta=0$, as expected, since canonical lifts are special covariant canonical transformations \cite{GIMMSY1, DeLeon1}.
Now, let us consider the classical field theory defined by the action
\eqref{APFT}, whose symmetry group $\mathcal{G}$ acts on its associated covariant configuration bundle $\fibbun{Y}{X}$ and transitively on its jet fibre-bundle $\fibbun{J^{1}\!Y}{Y}$, as previously described. For $\eta\in\mathcal{G}$, we denote by $\eta_{X}$, $\eta_{Y}$ and $\eta_{Z}$ their
corresponding transformations on $X$, $Y$ and $Z$, respectively. In a similar manner, given the infinitesimal generator $\xi_{\eta}\in\mathfrak{g}$ of $\eta$, we define $\xi_{\eta}^{X}$, $\xi_{\eta}^{Y}$  and $\xi_{\eta}^{Z}$ to be the vector fields associated to the transformations $\eta_{X}$, $\eta_{Y}$ and $\eta_{Z}$, correspondingly. Thus, we will say that $\mathcal{G}$ acts on $Z$ by covariant canonical transformations if the following condition holds
\begin{equation}\label{ICCT}
\mathfrak{L}_{\xi_{\eta}^{Z}}\Omega=0\, .
\end{equation}
Analogously, we say that $\mathcal{G}$ acts on $ Z$ by special covariant canonical transformations if the following identity is satisfied 
\begin{equation}\label{ISCCT}
\mathfrak{L}_{\xi_{\eta}^{Z}}\Theta=0 \, .
\end{equation}

We will further consider that, if $\mathcal{G}$ acts on $Z$ by covariant canonical transformations \eqref{ICCT}, the map from the multimomenta 
phase-space $Z$ to the space of linear mappings from the algebra $\mathfrak{g}$
of the Lie group $\mathcal{G}$ to  the vector space of $(n-1)$-forms on 
$Z$,
$\mathcal{J}^{(\mathcal{M})}: Z\rightarrow L\left(\mathfrak{g}, \Lambda^{n-1}Z\right)$, defined by
\begin{equation}\label{CMM}
d\,\mathcal{J}^{(\mathcal{M})}\!\left(\xi_{\eta}\right)=\xi_{\eta}^{Z}\lrcorner\,\Omega\, ,
\end{equation}
corresponds to the covariant momentum mapping associated to the action of $\mathcal{G}$ on the covariant phase-space $\left(Z,\Omega\right)$, where $\mathcal{J}^{(\mathcal{M})}\left(\xi_{\eta}\right)\in\Lambda^{n-1}Z$ is the $(n-1)$-form on $Z$ whose value at $z\in Z$ is given by $\left\langle\mathcal{J}^{(\mathcal{M})}(z), \xi_{\eta}\right\rangle$ for all $\xi_{\eta}\in\mathfrak{g}$, with $\xi_{\eta}^{Z}\in\mathfrak{X}\left(Z\right)$ denoting the vector field on $Z$ associated to $\xi_{\eta}$. By construction, $\mathcal{J}^{(\mathcal{M})}\!\left(\xi_{\eta}\right)$ is the analogous to the conserved current on $Z$ \cite{GIMMSY1, Crampin, DeLeon1}. As we will see below, the vanishing of those covariant momentum mappings on the space of Cauchy data will be related to the first class constraints, within Dirac's terminology, of a given classical field theory as discussed in~\cite{DeLeon1, GIMMSY2, Fischer}.

\subsection{Space plus time decomposition for classical field theory}\label{STDCFT}

In this subsection we will give a brief description of the space plus time decomposition of the covariant geometric Lagrangian and multisymplectic formalisms for classical field theory. To this end, we start by introducing the definition of a slicing of the base space $X$. Consequently, we describe the space plus time decomposition of an arbitrary fibre-bundle $\left(K, \fibbun{K}{X}, X\right)$ over $X$, with $K$ denoting the associated total space. This last will allow us to introduce the basis of the instantaneous Lagrangian and Hamiltonian formulations for classical field theory. Finally, we will relate the multisymplectic framework with the instantaneous Hamiltonian program through a reduction process. This section is based in references \cite{GIMMSY1, DeLeon1, DeLeon2, GIMMSY2, Fischer, Vankerschaver, Gotay1}.

To start the space plus time decomposition of the background  manifold $X$, let $\Sigma$ be a compact $(n-1)$-dimensional smooth manifold without boundary, such that, a slicing of $ X$ is defined as a diffeomorphism $\Psi:\mathbb{R}\times \Sigma\rightarrow  X$ \cite{GIMMSY2}. By taking into account that $\Psi_{t}:=\Psi\left(t,\cdot\right):\Sigma\rightarrow X$ defines an embedding, we introduce $\mathscr{X}:=\{\Psi_{t}\left| \right. t\in\mathbb{R}\}$ to be the set of all embeddings $\Psi_{t}$ of $\Sigma$ into $ X$ and we also introduce $\Sigma_{t}\subset X$ to represent the image of $\Sigma$ by $\Psi_{t}$, such that there exists $t_{0}\in\mathbb{R}$ satisfying $\Sigma_{t_{0}}=\Sigma$. 

Now, let $\left(t,u\right)$ be an arbitrary point of the space $\mathbb{R}\times\Sigma$. We define $\partial_{t}:=\partial/\partial t\in\mathfrak{X}\left(\mathbb{R}\times\Sigma\right)$ to be the generator of time translations. Then, the infinitesimal generator of the slicing $\Psi$ on $X$ is given by $\zeta_{ X}:=\Psi_{*}\left(\partial_{t}\right)\in\mathfrak{X}\left( X\right)$, which by construction is everywhere transverse to $\Sigma_{t}$. Thus, we will consider $\{x^{\mu}\}_{\mu=0}^{n-1}$ as a set of local coordinates adapted to the slicing on $ X$, where the Cauchy surfaces $\Sigma_{t}$ and the infinitesimal generator $\zeta_{X}$ are locally given by the level sets of the coordinate $x^{0}$ and $\partial_{0}$, respectively, as discused in \cite{DeLeon2, Vankerschaver}. 

As expected, the space plus time decomposition process will induce a similar decomposition for any geometric object defined on the base space $X$, which is the case of the fibre-bundles over it. To see this, let $\left(K, \fibbun{K}{X}, X\right)$ be a generic finite-dimensional fibre-bundle over $X$, where $K$ stands for the total space of the fibre-bundle. Given a slicing $\Psi$ of $X$ and a fibre-bundle $\left(K_{\Sigma},\fibbun{{\,K_{\Sigma}}}{\Sigma},\Sigma \right)$ over $\Sigma$, with $K_{\Sigma}$ denoting the associated total space, a compatible slicing of $K$ is defined by a bundle-diffeomorphism $\Phi:\mathbb{R}\times K_{\Sigma}\rightarrow K$, such that, the following diagram commutes
\[
\xymatrix{
\mathbb{R}\times K_{\Sigma} \ar[d] \ar[r]^{~~~\Phi} &  K \ar[d] \\
\mathbb{R}\times\Sigma \ar[r]^{~~~\Psi} &  X
}
\]
where the vertical arrows denote fibre-bundle projections. As before, we denote by $\Phi_{t}:=\Phi\left(t,\cdot\right): K_{\Sigma}\rightarrow K$ the embedding of $K_{\Sigma}$ by $\Phi$ into $K$ and we also introduce $K_{t}\subset K$ to denote the image of $K_{\Sigma}$ by $\Phi_{t}$. Additionally, we define $\zeta_{K}$ to be the infinitesimal generator of the slicing $\Phi$ of $K$, which is everywhere transverse to $K_{t}$. In particular, if $\zeta_{K}$ projects into $\zeta_{X}$ by means of the map $(\fibbun{K}{X})_\ast:TK\rightarrow TX$, we say that, the slicing $\left(\zeta_{K}, \zeta_{X}\right)$ of the fibre bundle $\fibbun{K}{X}$ is a compatible slicing, which defines a one-parametric group of bundle-automorphisms, that is, its associated vector flow is a fibre-preserving map \cite{GIMMSY2}. Here, we will consider $\left(x^{\mu}, k^{a}\right)$ as a set of local fibre-coordinates adapted to the slicing on $K$, such that, the surfaces $K_{t}$ are given by the level sets of $x^{0}$.

Now, let $\mathscr{K}$ be the set of all sections of the fibre-bundle $\fibbun{K}{X}$.  We denote by $\mathscr{K}_{t}$ the set of all sections of the restricted fibre-bundle $\fibbun{\,K_{t}}{\Sigma_{t}}:K_{t}\rightarrow\Sigma_{t}$, that is, the restriction of the fibre-bindle $\fibbun{K}{X}$ to the Cauchy surface $\Sigma_{t}$. Then, the collection $\mathscr{K}^{\Sigma}$ given by
\begin{equation}
\mathscr{K}^{\Sigma}:=\bigcup_{\Psi_{t}\in\mathscr{X}}\mathscr{K}_{t},
\end{equation}
defines an infinite-dimensional fibre-bundle over  the set of 
embeddings $\mathscr{X}$. Consequently, a section of the fibre-bundle $\fibbun{K}{X}$ induces a section of the fibre-bundle $\fibbun{\,\mathscr{K}^{\Sigma}}{\mathscr{X}}:\mathscr{K}^{\Sigma}\rightarrow \mathscr{X}$, and conversely.  This decomposition of the fibre-bundle will allow us to relate the finite- and infinite-dimensional descriptions of classical field theory \cite{DeLeon2}.

Besides, as $\mathscr{K}_{t}$ defines a smooth manifold, it is possible to construct its associated tangent, $T_{\sigma}\mathscr{K}_{t}$, and cotangent spaces, $T_{\sigma}^{*}\mathscr{K}_{t}$. 
The vector fields in $T_\sigma \mathscr{K}_t$ at $\sigma\in\mathscr{K}_t$ 
may be identified with the sections of $VK_t$ on $\sigma(\Sigma_t)\subset
K_t$, while the elements of the cotangent space $T_\sigma^\ast \mathscr{K}_t$ correspond to the linear mappings from the tangent space $T_\sigma \mathscr{K}_t$ to the $(n-1)$-forms in the Cauchy surface $\Sigma_t$, 
as described in~\cite{Forger2, Fischer}.
In particular, it is possible to induce differential forms on $\mathscr{K}_{t}$ from those defined on $ K$. To see this, let $\alpha$ be an $(s+n-1)$-form on $ K$, we define its associated differential form $\alpha_{t}$ on $\mathscr{K}_{t}$ as an $s$-form whose dual pairing with a set of $s$ tangent vectors $V_{i}\in T_{\sigma}\mathscr{K}$ at $\sigma\in\mathscr{K}_{t}$ explicitly reads
\begin{equation}\label{IFSS}
\alpha_{t}\left(\sigma\right)\left(V_{1},\cdots\!,V_{s}\right):=\int_{\Sigma_{t}}\sigma^{*}\left(i_{V_{1}\cdots V_{s}}\alpha\right)\, ,
\end{equation} 
where the contraction is taking along the image of $V_{i}\in T_{\sigma}\mathscr{K}_{t}$ \cite{Vankerschaver}. These definitions will allow us to introduce the bases of the infinite-dimensional description of classical field theory, as we will see below.

Let $\left(\zeta_{X}, \zeta_{Y}\right)$ be a compatible slicing of the covariant configuration bundle $\fibbun{Y}{X}$, which we assume leaves the action principle \eqref{APFT} invariant, that is, $\zeta_{Y}$ contains the information associated to the symmetry group of the theory $\mathcal{G}$, as discussed above. Now, given $\Sigma_{t}$ a Cauchy surface locally defined by the level sets of $x^{0}$, we introduce $\{x^{i}\}_{i=1}^{n-1}$ and $i_{t}:\Sigma_{t}\rightarrow X$ to be the coordinate set of $\Sigma_{t}$ and the inclusion map, respectively. We also define $Y_{t}:=Y\!\left|\right.\!_{\Sigma_{t}}$ as the restriction of the covariant configuration bundle to the Cauchy surface $\Sigma_{t}
$, which gives rise to the fibre bundle $ \fibbun{Y_{t}}{\Sigma_{t}}: Y_{t} \rightarrow \Sigma_{t}
$ whose local coordinates are given by $\left(x^{i},y^{a}\right)$ and its space of all sections denoted by $\mathscr{Y}_{t}$, such that, there exists $\phi\in\mathscr{Y}$ ( a section of the fibre-bundle $Y$) that induces an element $\varphi\in\mathscr{Y}_{t}$ (a section of $Y_{t}$) by means of the inclusion map, that is, $\varphi=\phi\circ i_{t}$ \cite{DeLeon2, Fischer, Gotay1}. Thus, the instantaneous configuration space of the generic field theory \eqref{APFT} at time $t$ is defined by $\mathscr{Y}_{t}$. As usual, we define $T\,\mathscr{Y}_{t}$ as the instantaneous space of velocities whose local coordinates are given by $\left(y^{a},\dot{y}^{a}\right)$, where the instantaneous Lagrangian formalism takes place with $\varphi\in\mathscr{Y}_{t}$ representing the fields of the theory and the temporal evolution of the fields variables defined by
\begin{equation}\label{TEFV}
\dot{\varphi}:=\mathfrak{L}_{\zeta_{Y}}\varphi=T\varphi\left(\zeta_{ X}\right)-\zeta_{Y}\left(\varphi\right)\, ,
\end{equation} 
where $\mathfrak{L}_{\zeta_{Y}}$ denotes the Lie derivative of $\varphi$ along $\zeta_{Y}$, while $\varphi_\ast:T\,\Sigma_{t}\rightarrow T_\varphi Y_{t}$ stands for the differential map of the section $\varphi$, as discussed in detail in \cite{GIMMSY2, Forger2}. Now, by defining $\left(J^{1}Y\right)_{t}$ as the restriction of the fibre-bundle $\fibbun{J^{1}\!Y}{X}:J^{1}Y\rightarrow X$ to the Cauchy surface $\Sigma_{t}$, whose local representation is given by $\left(x^{i},y^{a}, y^{a}_{\mu}\right)$, we introduce the affine bundle map $\beta_{\zeta
_{ X}}:\left(J^{1}Y\right)_{t}\rightarrow J^{1}\left(Y_{t}\right)\times VY_{t}$ over $Y_{t}$, which is locally defined by $\beta_{\zeta_{Y}}\left(x^{i},y^{a},y^{a}_{\mu}\right):=\left(x^{i}, y^{a}, y^{a}_{i}, \dot{y}^{a}\right)$. Thus, the map $\beta_{\zeta_{Y}}$ corresponds to the jet decomposition map and induces an isomorphism between $\left(j^{1}\mathscr{Y}\right)_{t}$ and $T\,\mathscr{Y}_{t}$ \cite{GIMMSY2}.

In particular, given $j^{1}\phi$ a section of the fibre-bundle $\fibbun{J^{1}\!Y}{X}$, we have that, $j^{1}\phi\circ i_{t}\in \left(j^{1}\mathscr{Y}\right)_{t}$, which together with the jet decomposition map $\beta_{\zeta_{Y}}$ allow us to define the instantaneous Lagrangian density, $\mathcal{L}_{t,\zeta_{Y}}:J^{1}\left(Y_{t}\right)\times VY_{t}\rightarrow \Lambda^{n-1}\Sigma_{t}$, of the theory by $\mathcal{L}_{t,\zeta_{Y}}\left(j^{1}\varphi, \dot{\varphi}\right):=\left(j^{1}\phi\circ i_{t}\right)^{*}\left(\zeta_ X\lrcorner\, \mathcal{L}\right)$, where $j^{1}\varphi$ corresponds to a section of the fibre-bundle $J^{1}\left(Y_{t}\right)\rightarrow \Sigma_{t}$, $\dot{\varphi}$ is an element of $VY_{t}$, $\mathcal{L}$ is the covariant Lagrangian density and we have used the isomorphism $\beta_{\zeta_{Y}}\left(j^{1}\phi\circ i_{t}\right)=\left(j^{1}\varphi,\dot{\varphi}\right)$. 
(Note that on the definition of the instantaneous Lagrangian density 
$\mathcal{L}_{t,\zeta_{Y}}$ above we used the fact that the Lagrangian density 
$\mathcal{L}$ is an $n$-form on $X$, and thus we  only have to consider 
the inner product with the generator $\zeta_X$.)
Consequently, we may find in an analogous manner the instantaneous Lagrangian, $L_{t,\zeta_{Y}}:T\,\mathscr{Y}_{t}\rightarrow\mathbb{R}$, explicitly given by
\begin{equation}\label{ILCFT}
L_{t,\zeta_{Y}}\left(\varphi,\dot{\varphi}\right):=\int_{\Sigma_{t}}L\left(j^{1}\varphi, \dot{\varphi}\right)\xi^{0}d^{n-1}x_{0}\, ,
\end{equation}
where $L$ stands for the covariant Lagrangian function and $\left(\varphi,\dot{\varphi}\right)\in T\, \mathscr{Y}_{t}$ and $d^{n-1}x_0$ 
denotes the $(n-1)$-volume form on the Cauchy surface 
$\Sigma_t$~\cite{GIMMSY2, Vankerschaver}. As usual, the instantaneous Legendre transformation is defined by
\begin{equation}
\begin{aligned}
\mathbb{F}L_{t,\zeta_{Y}}:~~T\,\mathscr{Y}_{t}~~&\rightarrow~~T^{*}\mathscr{Y}_{t}\\
\left(\varphi,\dot{\varphi}\right)&\mapsto\left(\varphi,\pi\right)\, ,
\end{aligned}
\end{equation}
where $\pi_{a}$ denotes the instantaneous momenta explicitly given by $\pi_{a}:=\partial L /\partial \dot{\varphi}^{a}$, which in turn allows us to define the instantaneous $t$-primary constraint set of the theory $\mathscr{P}_{t}\subset T^{*}\mathscr{Y}_{t}$, that is, the restriction of the cotangent bundle $T^{*}\mathscr{Y}_{t}$ over $\mathscr{Y}_{t}$ to the image of the instantaneous Legendre map \cite{DeLeon2, Gotay1}.

Now, we will proceed to decompose the multisymplectic phase-space and introduce a reduction process to relate the multisymplectic and the instantaneous Hamiltonian frameworks. To this end, let $Z_{t}:=Z\left|\right._{\Sigma_{t}}$ be the restriction of the fibre-bundle $\fibbun{Z}{X}:Z\rightarrow X$ to the Cauchy surface $\Sigma_{t}$, that gives rise to the fibre-bundle $Z_{t}\rightarrow\Sigma_{t}$ whose set of all sections will be denoted by $\mathscr{Z}_{t}$. In particular, by means of \eqref{IFSS}, the space $\mathscr{Z}_{t}$ is endowed with the forms $\Theta_{t}$ and $\Omega_{t}$ associated to the forms $\Theta$ and $\Omega$ on $Z$, respectively, which implies that $\left(\mathscr{Z}_{t}, \Omega_{t}\right)$ is a presymplectic space \cite{Gotay1}. Thus, we introduce the fibre-bundle map $R_{t}: \mathscr{Z}_{t}\rightarrow T^{*}\mathscr{Y}_{t}$ over $\mathscr{Y}_{t}$ defined by
\begin{equation}
\begin{aligned}
\left\langle R_{t}\left(\sigma\right),V \right\rangle:=\int_{\Sigma_{t}}\varphi^{*}\left(V\lrcorner\,\sigma\right)\, ,
\end{aligned}
\end{equation}
with $\varphi=\pi_{\!{Y\!Z}}\circ\sigma\in\mathscr{Y}_{t}$ and $V\in T_{\varphi}\mathscr{Y}_{t}$, and the contraction is taken along the images of $V$ and $\sigma\in\mathscr{Z}_{t}$, respectively, which in local coordinates can be written as $R_{t}\left(\sigma\right):=\left(p^{0}_{a}\circ\sigma\right)dy^{a}\otimes d^{n-1}x_{0}$ and whose kernel is given by $\mathrm{ker}\,R_{t}:=\{\sigma\in\mathscr{Z}_{t}\left|\right. p^{0}_{a}\circ\sigma=0\}$. Then, we have that, the quotient map $\mathscr{Z}_{t}/\mathrm{ker}\,R_{t}\rightarrow T^{*}\mathscr{Y}_{t}$  is a symplectic diffeomorphism \cite{DeLeon2, GIMMSY2, Gotay1}. In particular, given the subset $\mathscr{N}_{t}:=\{\sigma\in\mathscr{Z}_{t}\left|\right. \sigma=\mathbb{F}\mathcal{L}\circ j^{1}\phi\circ i_{t}\}$ and 
considering $\sigma\in \mathscr{N}_{t}$, it is possible to write
\begin{equation}
R_{t}\left(\sigma\right)=\frac{\partial L}{\partial y^{a}_{0}}\!\left(j^{1}\varphi, \dot{\varphi}\right) dy^{a}\otimes d^{n-1}x_{0}\, ,
\end{equation}
which corresponds to an element of the instantaneous primary constraint set $\mathscr{P}_{t}$, such that, a section $\sigma\in\mathscr{N}_{t}$ that projects onto $\left(\varphi,\pi\right)\in\mathscr{P}_{t}\subset T^\ast\mathscr{Y}$ by means of the map $R_{t}$ is called a holonomic lift of $\left(\varphi,\pi\right)$ \cite{Gotay1}.

With all this in mind, we are in the position to study the action of 
the Lie group $\mathcal{G}$, the symmetry group of the theory, on $\mathscr{Z}_{t}$ and $T^{*}\mathscr{Y}_{t}$. Since $\mathcal{G}$ acts on $Z$ by covariant canonical transformations, the covariant momentum map $\mathcal{J}^{(\mathcal{M})}:Z\rightarrow L\left(\mathfrak{g},\Lambda^{n-1}Z\right)$ associated to the action of $\mathcal{G}$ induces, by means of \eqref{IFSS}, the so-called energy-momentum map $E_{t}:\mathscr{Z}_{t}\rightarrow g^{*}$ that explicitly is given by
\begin{equation}
\label{eq:energy}
\left\langle E_{t}\left(\sigma\right),\xi \right\rangle:=\int_{\Sigma_{t}}\sigma^{*}\mathcal{J}^{(\mathcal{M})}\!\left(\xi\right)\, ,
\end{equation}
for each $\xi\in\mathfrak{g}$. Now, we define $\mathcal{G}_{t}:=\{\eta\in\mathcal{G}\left|\right. \eta_{X}\left(\Sigma_{t}\right)=\Sigma_{t}\}$ to be the subset of $\mathcal{G}$ that acts on $\Sigma_{t}$ by diffeomorphisms. Thus, $\mathcal{G}_{t}$ is a canonical action on $\mathscr{Z}_{t}$ which allows us to identify, for each $\eta\in\mathcal{G}_{t}$, the maps $\eta_{t}:=\eta_{X}\left|\right._{\Sigma_{t}}$ as elements of the group of diffeomorphisms on $\Sigma_t$, $\mathrm{Diff}\left(\Sigma_{t}\right)$.  The infinitesimal generators of $\mathrm{Diff}\left(\Sigma_{t}\right)$ will be denoted by $\xi_{\eta_{t}}^{X}$, which in adapted coordinates satisfy the condition $\xi^{0}(x)=0$ on $\Sigma_{t}$, that is, the corresponding vector field $\xi_{\eta_{t}}^{X}$ on $X$ associated to $\xi_{\eta_{t}}$ is tangent to the 
Cauchy surface $\Sigma_{t}$. In consequence, the energy-momentum map $E_{t}$ restricted to the subspace $\mathcal{G}_{t}\subset\mathcal{G}$ gives rise to the map $\mathcal{J}_{t}:=E_{t}\left|\right._{\mathfrak{g}_{t}}:\mathscr{Z}_{t}\rightarrow \mathfrak{g}_{t}^{*}$, which corresponds to a momentum map for the action of $\mathcal{G}_{t}$ on $\mathscr{Z}_{t}$, as discussed in detail in \cite{DeLeon2, GIMMSY2, Fischer}. In particular, since $\left(T^{*}\mathscr{Y}_{t},\omega_{t}\right)$ is the symplectic quotient of the presymplectic space $\left(\mathscr{Z}_{t}, \Omega_{t}\right)$ by the kernel of the map $R_{t}$, we have that, the momentum map $\mathscr{J}_{t}:T^{*}\mathscr{Y}_{t}\rightarrow \mathfrak{g}_{t}^{*}$ associated to the action of $\mathcal{G}_{t}$ on $T^{*}\mathscr{Y}_{t}$ is given by
\begin{equation}\label{PMM}
\left\langle \mathscr{J}_{t}\left(\varphi,\pi\right),\xi \right\rangle:=\left\langle \mathcal{J}_{t}\left(\sigma\right),\xi \right\rangle\, ,
\end{equation}
where $\xi\in\mathfrak{g}_{t}$, $\left(\varphi,\pi\right)\in T^{*}\mathscr{Y}_{t}$ and $\sigma$ is an element of $R_{t}^{-1}\{\varphi,\pi\}\subset \mathscr{Z}_{t}$. In this way, definition~\eqref{PMM} only reflects 
the restriction of the energy-momentum map~\eqref{eq:energy} to the 
subspace  $\mathcal{G}_{t}\subset\mathcal{G}$.

Besides, since the canonical prolongation of $\zeta_{Y}$ to $J^{1}Y$ leaves the action~\eqref{APFT} invariant, the corresponding canonical lift of $\zeta_{Y}$ to $Z$, $\zeta_{Z}\in \mathfrak{X}\left(Z\right)$, acts on $Z$ by covariant canonical transformations. Then, $\zeta_{Z}$ has an associated covariant momentum map, which projects onto a well defined function on the $t$-primary constraint set $\mathscr{P}_{t}\in T^{*}\mathscr{Y}_{t}$, which coincides with the instantaneous Hamiltonian of the theory, $H_{t,\zeta_{Y}}:\mathscr{P}_{t}\rightarrow \mathbb{R}$, and can be explicitly defined by
\begin{equation}\label{IHT}
H_{t,\zeta_{Z}}\!\left(\varphi,\pi\right):=-\int_{\Sigma_{t}}\sigma^{*}\mathcal{J}^{(\mathcal{M})}\!\left(\zeta_{Z}\right)\, ,
\end{equation}
being $\sigma$ the holonomic lift associated to $\left(\varphi,\pi\right)\in\mathscr{P}_{t}$, that is, $\sigma\in\mathscr{N}_{t}$ \cite{DeLeon2, GIMMSY2, Fischer}. 

Furthermore, by extending the second Noether theorem to the multisymplectic approach~\cite{GIMMSY1, GIMMSY2, Fischer}, we may see that the projected momentum map $\left\langle \mathscr{J}_{t}\left(\varphi,\pi\right),\xi \right\rangle$ on $T^{*}\mathscr{Y}_{t}$ corresponds to the generator of the infinitesimal gauge transformations of the theory. In particular,  the first class constraints of the theory, within Dirac's terminology \cite{QGS}, are encoded on the zero level set of the projected momentum map $\left\langle \mathscr{J}_{t}\left(\varphi,\pi\right),\xi \right\rangle$, that is, the subspace on $T^{*}\mathscr{Y}_{t}$ defined by $\mathscr{J}_{t}^{-1}\left(0\right):=\{\left(\varphi,\pi\right)\in T^{*}\mathscr{Y}_{t}\left|\right. \left\langle \mathscr{J}_{t}\left(\varphi,\pi\right),\xi \right\rangle=0\}$ corresponds to the surface on the instantaneous phase-space, $T^{*}\mathscr{Y}_{t}$, characterized by the first class constraints of the system, which arise in the instantaneous Dirac-Hamiltonian analysis of the theory \cite{QGS}.

\section{Non-Abelian topological BF theory}
\label{sec:BFTheory}

We start this section by giving a brief description of the non-Abelian topological BF theory on a $4$-dimensional space-time manifold. Then, we proceed to describe the features of this topological field theory within the Lagrangian and the multisymplectic geometric approaches for classical field theory, focusing our attention on the study of the symmetries of the model.  In particular, these symmetries will play a fundamental role in order to relate the multisymplectic and the instantaneous Dirac-Hamiltonian formulations of the BF theory, which will be discussed in detail at the end of the section.

We will start by introducing the model of our interest, which is a $4$-dimensional non-Abelian BF theory. BF theories are a class of topological diffeomorphism invariant field theories.  As it is well known, they are non-metric field theories and have no local degrees of freedom (see for instance \cite{Montesinos1, Escalante1, Sardanashvily1, Cattaneo, Horowitz}). BF theories have a strong relationship with Einstein's theory of General Relativity since it is possible to write the latter as a constrained BF theory giving rise to the so-called BF gravity models 
(see~\cite{Montesinos2}, and references therein, for details), which are based on the Plebanski formulation of General Relativity developed in \cite{Plebanski}. Thus, BF theories are very interesting from the physical point of view and have been widely explored from different prespectives and approaches at classical and quantum level, as we can found in the literature (see for instance \cite{Cattaneo, Horowitz, Montesinos2, Spinfoams1, Spinfoams2, Angel}). In particular, we are interested in analysing a non-Abelian BF theory within the previously described multisymplectic formalism for classical field theory.
We will closely follow the description of the model developed in \cite{Cattaneo}

Now, we proceed to describe the 4-dimensional non-Abelian BF theory. In order to do that, let $X$ be a $4$-dimensional space-time manifold without boundary endowed with Minkowski signature $\mathrm{diag}(-1,+1,+1,+1)$ and locally represented by $\{x^{\mu}\}^{3}_{\mu=0}$. Let us consider $\mathcal{G}$ a compact simple Lie group that will be the gauge symmetry group of the theory. The non-Abelian BF theory is defined by the action principle
\begin{equation}\label{BFT}
S_{\mathrm{BF}}\left[A ^{a},B^{a}\right]:=\int_{X}B^{a}\wedge F_{a}\, ,
\end{equation}
where $F^{a}:=dA^{a}+\frac{1}{2}\left[A\wedge A\right]^{a}$ denotes the curvature 2-form associated to the connection 1-form valuated on the Lie algebra 
$\mathfrak{g}$, $A^{a}\in\Lambda^{1}X$, while $B^{a}\in\Lambda^{2}X$ corresponds to a set of $\mathfrak{g}$-valued $2$-forms, $d$ stands for the exterior derivative on $X$ and $[\cdot\wedge\cdot]:\Lambda^{q}X\times \Lambda^{p}X\rightarrow \Lambda^{q+p}X$ is the graded commutator for $\mathfrak{g}$-valued differential forms which, given the $\mathfrak{g}$-valued forms $\alpha\in \Lambda^{q}X$ and $\beta\in\Lambda^{p}X$, explicitly reads as $[\alpha\wedge\beta]:=\alpha\wedge\beta-(1)^{pq}\beta\wedge\alpha$ \cite{Cattaneo}.

As previously mentioned,  a BF theory is a diffeomorphism invariant field theory, but this type of symmetry is not the unique symmetry of the model. In fact, BF theory is also invariant with respect to the gauge transformations generated by the action of the gauge symmetry group $\mathcal{G}$ and the transformations associated to the so-called topological symmetry. However, these symmetries are not all independent, since it is possible to build the infinitesimal diffeomorphism transformations on the field variables by means of the gauge and topological symmetries of the theory, which are associated with the first class constraints that arises within the Dirac-Hamiltonian formulation of the model, as discussed in \cite{Montesinos1, Escalante1, Montesinos3}. With this in mind, we will focus our attention to the analysis of the 
gauge and the topological symmetries of the non-Abelian BF theory. 
On the one hand, the infinitesimal transformations associated to the gauge symmetry group of the BF model read
\begin{equation}\label{GSBF}
\begin{aligned}
A^{a}&\rightarrow A^{a}_{\theta}:=A^{a}+d_{A}\theta^{a}\, ,\\
B^{a}&\rightarrow B^{a}_{\theta}:=B^{a}+[B\wedge\theta]^{a}\, ,
\end{aligned}
\end{equation}
where $d_{A}:\Lambda^{q}X\rightarrow\Lambda^{q+1}X$ denotes the covariant exterior derivative explicitly defined as $d_{A}:=d+[A\wedge \cdot]$ and $\theta^{a}\in\Lambda^{0}X$ is a set of $\mathfrak{g}$-valued functions on $X$. It is possible to see that, under these infinitesimal transformations, $A^{a}$ transforms as a connection, while $B^{a}$ transforms as a covariant 2-tensor, bringing the local expressions
\begin{equation}\label{LEGT}
\begin{aligned}
{A_{\theta}}^{a}_{\mu}&=A^{a}_{\mu}+D_{\mu}\theta^{a}\, ,\\
{B_{\theta}}^{a}_{\mu\nu}&=B^{a}_{\mu\nu}+{f^{a}}_{bc}B^{b}_{\mu\nu}\theta^{c}\, ,
\end{aligned}
\end{equation}
where $D_{\mu}\theta^{a}:=\partial_{\mu}\theta^{a}+{f^{a}}_{bc}A^{a}_{\mu}\theta^{c}$ denotes the covariant derivative associated to the $1$-form $A^a$. On the other hand, the infinitesimal transformations associated to the topological symmetry of the BF theory are given by
\begin{equation}\label{TSBF}
\begin{aligned}
A^{a}&\rightarrow A^{a}_{\chi}:=A^{a}\, ,\\
B^{a}&\rightarrow B^{a}_{\chi}:=B^{a}+d_{A}\chi^{a}\, ,
\end{aligned}
\end{equation}
where $\chi^{a}\in\Lambda^{1}X$ denotes a set of $\mathfrak{g}$-valued $1$-forms on $X$. With respect to the topological symmetry, we see that $A^{a}$ remains unchanged while $B^{a}$ transforms as a connection~\cite{Cattaneo, Montesinos1}. Then, the topological infinitesimal transformations locally read
\begin{equation}\label{LETT}
\begin{aligned}
{A_{\chi}}^{a}_{\mu}&=0\, ,\\
{B_{\chi}}^{a}_{\mu\nu}&=B^{a}_{\mu\nu}+2D_{[\mu}\chi^{a}_{\nu]} \, ,
\end{aligned}
\end{equation}
where $\chi^{a}_{\mu}$ denotes the components of the $\mathfrak{g}$-valued $1$-forms $\chi^{a}$ and $D_{[\mu}\chi^{a}_{\nu]}:=\frac{1}{2}\left(D_{\mu}\chi^{a}_{\nu}-D_{\nu}\chi^{a}_{\mu}\right)$ stands for the anti-symmetrization in Greek indices. Bearing this in mind, in the next subsection we will analyze the non-Abelian topological BF theory \eqref{BFT} within the geometric Lagrangian approach putting special attention on the study of the symmetries of the model. 

\subsection{Geometric Lagrangian analysis}

Now, we will proceed to develop the geometric Lagrangian analysis of the non-Abelian topological BF theory focusing our attention on the study of the symmetries of the system. In particular, we want to construct the Noether currents associated to the BF theory which, as we will see in the following subsections,  will be strongly related to the covariant momentum maps of the model at the multisymplectic level. 

As it is well known, the topological BF theory on a $4$-dimensional space-time background manifold, $X$, corresponds to a classical field theory whose dynamical fields can be understood as sections of the vector fibre-bundle $\fibbun{Y}{X}: Y:= \overset{1}{\wedge}\,T^{*}X\oplus \overset{2}{\wedge}\,T^{*}X\rightarrow X$, whose local representation, at the point $y\in Y$, is given by $y:=(x^{\mu}, a_{\mu}^{a}, b^{a}_{\mu\nu})$ \cite{Sardanashvily1}. Thus, the first jet manifold associated to the covariant configuration space of the theory, $J^{1}Y$, is locally represented by $\gamma:=\left(x^{\mu},a^{a}_{\mu},b^{a}_{\mu\nu}, a^{a}_{\mu\nu}, b^{a}_{\mu\nu\sigma}\right)$ at the point $\gamma\in J^{1}Y$. Then, given $\mathscr{Y}$, the space of all smooth sections of the fibre-bundle $\fibbun{Y}{X}$, a section $\phi\in\mathscr{Y}$ at the point $x\in X$ locally reads as $\phi^{*}y=(x^{\mu},A_{\mu}^{a},B_{\mu\nu}^{a})$, such that, the section $j^{1}\phi$, the jet prolongation associated to $\phi$, of the fibre-bundle $\fibbun{J^{1}\!Y}{X}:J^{1}Y\rightarrow X$ at $x\in X$ can be locally represented as
\begin{equation}
j^{1}\phi^{*}\gamma=(x^{\mu}, A^{a}_{\mu}, B^{a}_{\mu\nu}, \partial_{\mu}A^{a}_{\nu}, \partial_{\mu}B^{a}_{\nu\sigma})\, ,
\end{equation}
where Greek indices label space-time coordinates while Latin indices take values on the Lie algebra $\mathfrak{g}$ of the gauge symmetry group $\mathcal{G}$ of the theory, respectively. In particular, since the fields $B^{a}_{\mu\nu}$ correspond to the components of a set of $\mathfrak{g}$-valued $2$-forms on $X$, they are completely antisymmetric in the Greek indices, that is, the condition $ B^{a}_{\mu\nu}=B^{a}_{[\mu\nu]}$ holds, implying that the number of linear independent field variables of the theory is $N\times\left(4+6\right)$, where $N$ denotes the number of the generators of the Lie algebra $\mathfrak{g}$, while  the 4 and the 6 stand for the linear independent components of the connection $A^a$ and the 2-forms $B^a$, respectively. Thus, in what follows, we will restrict our analysis to the linearly independent field variables of the model which will play an important role, as we will see below, in the space plus time decomposition of the multisymplectic formulation of the BF theory.

Now, we introduce the Lagrangian density of the system,  $\mathcal{L}_{BF}:J^{1}Y\rightarrow \Lambda^{4}X$, that from the action principle \eqref{BFT} can be locally defined as
\begin{equation}\label{LDBFT}
\mathcal{L}_{\mathrm{BF}}\left(\gamma\right):=\frac{1}{4}\epsilon^{\mu\nu\sigma\rho}b^{a}_{\mu\nu}F_{a\sigma\rho}\, d^{\,4}x\, ,
\end{equation} 
where $F^{a}_{\mu\nu}=a^{a}_{\mu\nu}-a^{a}_{\nu\mu}+{f^{a}}_{bc}a^{b}_{\mu}a^{c}_{\nu}$ stands for the components of the curvature $2$-form $F^{a}$ and $f_{abc}$ corresponds to the structure constants of the Lie algebra $\mathfrak{g}$, while $\epsilon^{\mu\nu\sigma\rho}$ denotes the Levi-Civita alternating symbol. Thus, from the definition \eqref{PCF}, the Poincaré-Cartan $4$-form of the BF  theory, $\Theta^{(\mathcal{L})}_{\mathrm{BF}}\in \Lambda^{4}J^{1}Y$, explicitly reads as  
\begin{equation}
\Theta^{(\mathcal{L})}_{\mathrm{BF}}:=\frac{1}{2}\epsilon^{\mu\nu\sigma\rho}\,b_{a\sigma\rho}\left(da^{a}_{\nu}\wedge d^{3}x_{\mu}+\frac{1}{2}{f^{a}}_{bc}a^{b}_{\mu}a^{c}_{\nu}\,d^{4}x\right)\,.
\end{equation}
As stated in the previous section, this will be the main geometric object to describe the symmetries of the BF theory. 

As previously mentioned, we are interested in analysing the symmetries of the BF theory within the geometric Lagrangian approach. To this end, we start by introducing the infinitesimal generators of the gauge~\eqref{LEGT} and the  topological~\eqref{LETT} transformations of the BF theory, which are elements of the vector space $\mathfrak{X}\left(Y\right)$ and are locally defined by
\begin{subequations}\label{IGGTS}
\begin{align}
\xi_{\theta}^{Y}&:=D_{\mu}\theta^{a}\frac{\partial}{\partial a^{a}_{\mu}}+\frac{1}{2!} {f^{a}}_{bc}b^{b}_{\mu\nu}\theta^{c}\frac{\partial}{\partial b^{a}_{\mu\nu}}\, ,\label{IGGS}\\
\xi_{\chi}^{Y}&:=D_{[\mu}\chi^{a}_{\nu]}\frac{\partial}{\partial b^{a}_{\mu\nu}}\, , \label{IGTS}
\end{align}
\end{subequations} 
where we are taking $D_{\mu}\theta^{a}=\partial_{\mu}\theta^{a}+{f^{a}}_{bc}a^{b}_{\mu}\theta^{a}$ and we have introduced the factor $1/2!$ in order to consider the antisymmetry properties of the variables $b^{a}_{\mu\nu}$ and we have also restricted our analysis to only consider the linearly independent variables of the theory\footnote{In particular, by taking into account the antisymmetry of the variables $b^{a}_{\nu\sigma}=b^{a}_{[\nu\sigma]}$ and $b^{a}_{\mu\nu\sigma}=b^{a}_{\mu[\nu\sigma]}$, we will follow the conventions $\displaystyle{\frac{\partial b^{a}_{\nu\sigma}}{\partial b^{b}_{\beta\gamma}}=\delta^{a}_{b}\delta^{\beta\gamma}_{\nu\sigma}}$ and $\displaystyle{\frac{\partial b^{a}_{\mu\nu\sigma}}{\partial b^{b}_{\alpha\beta\gamma}}=\delta^{a}_{b}\delta^{\alpha}_{\mu}\delta^{\beta\gamma}_{\nu\sigma}}$, where $\delta^{\beta\gamma}_{\nu\sigma}$ denotes the generalized Kronecker delta.}.

We know that the action of the symmetry group of a classical field theory on the associated covariant configuration space induces transformations on the jet fibre-bundle \cite{DeLeon1, GIMMSY2}, which implies that the canonical lifts \eqref{PJB} associated to the vector fields \eqref{IGGTS} correspond to the generators of the infinitesimal gauge and topological transformations on the jet bundle $\fibbun{J^{1}\!Y}{Y}$ associated to the BF theory.  Those lifts are thus elements of the vector space $\mathfrak{X}\left(J^{1}Y\right)$ and can be explicitly written as
\beq
\xi_{\theta}^{J^{1}\!Y}
&:= & 
\!D_{\mu}\theta^{a}\frac{\partial}{\partial a^{a}_{\mu}}+\frac{1}{2!}{f^{a}}_{bc}b^{b}_{\mu\nu}\theta^{c}\frac{\partial}{\partial b^{a}_{\mu\nu}}+\left(D_{\nu}\left(\partial_{\mu}\theta^{a}\right)\!+\!{f^{a}}_{bc}a^{b}_{\mu\nu}\theta^{c}\right)\frac{\partial}{\partial a^{a}_{\mu\nu}}
\nn\\
& & 
+\frac{1}{2!}{f^{a}}_{bc}\left(b^{b}_{\nu\sigma}\partial_{\mu}\theta^{c}\!+\!b^{b}_{\mu\nu\sigma}\theta^{c}\right)\frac{\partial}{\partial b^{a}_{\mu\nu\sigma}}\,,
\nn\\
\xi_{\chi}^{J^{1}\!Y}
&:= & 
D_{[\mu}\chi^{a}_{\nu]}\frac{\partial}{\partial b^{a}_{\mu\nu}}+\left(D_{[\nu}\left(\partial_{\mu}\chi^{a}_{|\sigma]}\right)+{f^{a}}_{bc}a^{b}_{\mu[\nu}\chi^{c}_{\sigma]}\right)\frac{\partial}{\partial b^{a}_{\mu\nu\sigma}}\, ,
\eeq
where as before the factor $1/2!$ has been introduced to consider the antisymmetry of the variables $b^{a}_{\mu\nu}$ and $b^{a}_{\mu\nu\sigma}=b^{a}_{\mu[\nu\sigma]}$, thus restricting our analysis to the linearly independent variables of the theory. 

On the one hand, a direct calculation shows that the gauge symmetry \eqref{GSBF} corresponds to a natural symmetry~\eqref{eq:natural} of the BF theory, since the vector field $\xi_{\theta}^{J^{1}Y}\in\mathfrak{X}\left(J ^{1}Y\right)$ satisfies the condition
\begin{equation}
\label{eq:BFnat}
\mathfrak{L}_{\xi_{\theta}^{J^{1}\!Y}}\Theta^{(\mathcal{L})}_{\mathrm{BF}}=0\, ,
\end{equation} 
where $\mathfrak{L}$ denotes the Lie derivative defined on $J^{1}Y$, 
therefore implying that the Lagrangian density \eqref{LDBFT} is equivariant with respect to the infinitesimal transformations associated to the action of the gauge symmetry group of the theory, while, on the other hand, the topological symmetry \eqref{TSBF} corresponds to a Noether 
symmetry~\eqref{eq:noethersym} of the BF theory, since the vector field $\xi_{\chi}^{J^{1}Y}\in\mathfrak{X}\left(J ^{1}Y\right)$ satisfies the relation
\begin{equation}
\label{eq:BFNoether}
\mathfrak{L}_{\xi_{\chi}^{J^{1}\!Y}}\Theta^{(\mathcal{L})}_{\mathrm{BF}}=d\alpha\, ,
\end{equation}
where $\alpha\in\Lambda^{3}\left(J^{1}Y\right)$ is a $3$-form on $J^{1}\mathcal{Y}$ explicitly defined by
\begin{equation}\label{DFAlpha}
\alpha:=\frac{1}{2}\epsilon^{\mu\nu\sigma\rho}\chi_{a\nu}\left(da^{a}_{\rho}\wedge d^{2}x_{\mu\sigma}+{f^{a}}_{bc}a^{b}_{\sigma} a^{c}_{\rho}\,d^{3}x_{\mu}\right)\, .
\end{equation}
Here we have introduced the short notation $d^{2}x_{\mu\sigma}:=\partial_\sigma\lrcorner\partial_\mu\lrcorner
d^4 x$. 

Before proceeding to construct the Noether currents associated to the BF theory, we will obtain the field equations of the this topological field theory following the procedure discussed in subsection \eqref{Lagsec}. To this end, let $W\in\mathfrak{X}\left(J^{1}Y\right)$ be an arbitrary projectable vector field on $J^{1}Y$ locally defined at $\gamma\in J^{1}Y$ by
\begin{equation}
W:=W\left(x\right)^{\mu}\frac{\partial}{\partial x^{\mu}}+W(y)^{a}_{\nu}\frac{\partial}{\partial a^{a}_{\nu}}+\frac{1}{2!}W(y)^{a}_{\nu\sigma}\frac{\partial}{\partial b^{a}_{\nu\sigma}}+W(\gamma)^{a}_{\mu\nu}\frac{\partial}{\partial a^{a}_{\mu\nu}}+\frac{1}{2!}W(\gamma)^{a}_{\mu\nu\sigma}\frac{\partial}{\partial b^{a}_{\mu\nu\sigma}}\, ,
\end{equation}
where we have considered  $x=\fibbun{J^{1}\!Y}{X}\left(\gamma\right)\in X$ and $y=\fibbun{J^{1}\!Y}{Y}\left(\gamma\right)\in Y$ as elements of their 
respective parts of $J^{1}Y$. Then, given $W$ and considering $\phi\in\mathscr{Y}$ as a critical point of the action principle \eqref{APFT}, the condition $\left(j ^{1}\phi\right)^{*}\left(W\lrcorner\,\Omega^{(\mathcal{L})}_{\mathrm{BF}}\right)=0$ holds, giving rise to  relation
\beq
0 &= &
\frac{1}{2}\epsilon^{\mu\nu\sigma\rho}\left[-W^{\alpha}\left(x\right)\left(\frac{1}{2}\,\partial_{\alpha}B_{a\mu\nu}F^{a}_{\sigma\rho}\!+\!\partial_{\alpha}A^{a}_{\mu}D_{\nu}B_{a\sigma\rho}\right)
\right.
\nn\\
& & 
+\left.
\frac{1}{2}\,W_{a\mu\nu}\left(x,A,B\right)F ^{a}_{\sigma\rho}
+ W^{a}_{\mu}\left(x,A,B\right)D_{\nu}B_{a\sigma\rho}\right]
\, ,
\eeq
which by taking into account that $W$ is an arbitrary vector field on $J^{1}Y$, and by linear independence of the components of $W$, it is possible to obtain the equations
\beq
\label{BFFE}
\hspace{8ex} F^{a}_{\mu\nu}
& = & 
0\, ,\nn\\
\epsilon^{\mu\nu\sigma\rho}D_{\nu}B^{a}_{\sigma\rho}
& = & 0\, ,
\eeq
relations that correspond to the  field equations associated to the BF theory \cite{Montesinos3}.

We finish this subsection by computing the associated Noether currents of the BF theory. To do that, let $\phi\in\mathscr{Y}$ be a solution of the Euler-Lagrange field equations \eqref{BFFE}. Then, the $3$-forms on $X$ defined by
\begin{subequations}\label{NCBFT}
\begin{align}
{\mathcal{J}}^{(\mathcal{L})}_{\mathrm{BF}}(\theta)&:=\left(j^{1}\phi\right)^{*}\left(\xi_{\theta}^{J^{1}\!Y}\lrcorner\Theta^{(\mathcal{L})}_{\mathrm{BF}}\right)=\frac{1}{2}\epsilon^{\mu\nu\sigma\rho} D_{\nu}\theta^{a}B_{a\sigma\rho}d^{3}x_{\mu}\, ,\label{NCBFT1}\\
{\mathcal{J}}^{(\mathcal{L})}_{\mathrm{BF}}(\chi)&:=\left(j^{1}\phi\right)^{*}\left(\xi_{\chi}^{J^{1}\!Y}\lrcorner\Theta^{(\mathcal{L})}_{\mathrm{BF}}-\alpha\right)=-\frac{1}{2}\epsilon^{\mu\nu\sigma\rho}\chi_{a\nu}F^{a}_{\sigma\rho}d^{3}x_{\mu}\, ,\label{NCBFT2}
\end{align}
\end{subequations}
correspond to the conserved Noether currents obtained for this topological field theory by means of the Noether's theorem~\cite{Montesinos1}. 
It is possible to see that, on the space of solutions to the field 
equations~\eqref{BFFE},  the Noether  current associated to the gauge symmetry group of the theory~\eqref{NCBFT1} integrated over a Cauchy surface vanishes by requiring that gauge arbitrary functions on space-time to be of compact support on the boundary of the Cauchy surface, while  
the Noether current associated to the topological symmetry~\eqref{NCBFT2} is trivially zero on each point of the space-time manifold~\cite{Montesinos3}. 
As we will see below, the classification of the symmetries of the theory in 
natural~\eqref{eq:BFnat} and Noether~\eqref{eq:BFNoether} symmetries will play an important role in the multisymplectic formulation of the BF theory.
We will discuss this in detail in the following subsections.

\subsection{Multisymplectic analysis}
\label{sec:multiBF}

In the present subsection we will develop the multisymplectic analysis of the BF theory, focusing our attention on the study of the symmetries of the theory within this covariant and geometric approach. In order to do so, we start by constructing the multisymplectic phase-space of the theory and the relevant geometric objects to describe the features of  topological BF field theory. To this end, following the brief description of the multisymplectic formulation for classical field theory presented in  subsections~\ref{sec:multi} and~\ref{SCMPS}, we define the multimomenta phase-space of the BF theory as the subbundle over $Y$ defined by 
$\fibbun{Z}{X}:Z\subset \Lambda^{4}Y\rightarrow Y$, whose local coordinates are given by $(x^{\mu}, a^{a}_{\nu}, b^{a}_{\nu\sigma}, p, p_{a}^{\mu\nu}, p_{a}^{\mu\nu\sigma})$, where the multimomenta $p^{\mu\nu\sigma}_{a}$ associated to the variables $b^{a}_{\nu\sigma}$ are subject to the condition $p^{\mu\nu\sigma}_{a}=p^{\mu[\nu\sigma]}_{a}$, implying that, the number of linearly independent multimomenta variables of the theory is 
$4\times(N\times\left(4+6\right))$, where the extra 4 is due to the fact that we are considering the gradient in each of the space-time directions in the 
definition of the multimomenta. Thus, the canonical 4-form on $Z$, $\Theta_{\mathrm{BF}}\in\Lambda^{4}Z$,  defined by
\begin{equation}
\Theta_{\mathrm{BF}}:=p_{a}^{\mu\nu}da^{a}_{\nu}\wedge d^{3}x_{\mu}+\frac{1}{2!}\ p_{a}^{\mu\nu\sigma}db^{a}_{\nu\sigma}\wedge d^{3}x_{\mu}+p\,d^{4}x\, ,
\end{equation}
corresponds to the multisymplectic potential, such that the negative of its exterior derivative induces on $Z$ the multisymplectic form of the BF theory, $\Omega_\mathrm{BF}=-d\Theta_{\mathrm{BF}}\in\Lambda^{5}Z$, which is explicitly given by
\begin{equation}
\Omega_{\mathrm{BF}}:=da^{a}_{\nu}\wedge dp^{\mu\nu}_{a}\wedge d^{3}x_{\mu}+\frac{1}{2!}\ db^{a}_{\nu\sigma}\wedge dp^{\mu\nu\sigma}_{a}\wedge d^{3}x_{\mu}-dp\wedge d^{4}x\, ,
\end{equation}
where once again the factor $1/2!$ has been introduced in the above equations in order to consider the antisymmetry properties of the variables $b^{a}_{\nu\sigma}$ and $p^{\mu\nu\sigma}_{a}$, thus restricting  our analysis to the linearly independent variables of the theory\footnote{Similarly, by taking into account the antisymmetry of the multimomenta variables $p^{\mu\nu\sigma}_{a}=p^{\mu[\nu\sigma]}_{a}$, we will follow the convention $\displaystyle{\frac{\partial p^{\mu\nu\sigma}_{a}}{\partial p^{\alpha\beta\gamma}_{b}}=\delta^{b}_{a}\delta^{\mu}_{\alpha}\delta^{\nu\sigma}_{\beta\gamma}}$.}. Then, the pair $(Z,\Omega_{\mathrm{BF}})$ corresponds to the multisymplectic phase-space of the BF theory.

Now, we will study the symmetries of the BF theory within the multisymplectic approach. 
To this end, we will apply the theory of Lie groups acting on the multisymplectic phase-space as
described in subsection \eqref{SCMPS}. We will start by considering that, since the gauge symmetry \eqref{GSBF} corresponds to a natural symmetry of the theory~\eqref{eq:natural}, then the lift of the vector field \eqref{IGGS} from $TY$ to $TZ$ will generate special covariant canonical transformations on $Z$, that is, it will satisfy the condition \eqref{ISCCT} for the BF theory. However, the case of the topological 
symmetry~\eqref{TSBF} is different, since this last corresponds to a divergence 
symmetry~\eqref{eq:noethersym} of the theory.
Thus, we will expect that, in order to be consistent with the geometric Lagrangian analysis of the non-Abelian BF model, the topological symmetry acts on $Z$ producing  covariant canonical 
transformations~\eqref{ICCT}.
Indeed, this is the case as relation \eqref{CLMMPS}  brings the vector field $\xi_{\theta}^{Z}\in\mathfrak{X}\left(Z\right)$ associated to the action of the gauge symmetry group of the BF theory on $Z$, namely,
\beq
\label{GGSMPS} 
\hspace{-8ex}
\xi_{\theta}^{Z}
& := & 
D_{\mu}\theta^{a}\frac{\partial}{\partial a^{a}_{\mu}}+\frac{1}{2!}{f^{a}}_{bc}b^{b}_{\mu\nu}\theta^{c}\frac{\partial}{\partial b^{a}_{\mu\nu}}-\left(D_{\nu}(\partial_{\mu}\theta^{a})p^{\mu\nu}_{a}+\frac{1}{2!}{f^{a}}_{bc}b^{b}_{\nu\sigma}\partial_{\mu}\theta^{c}p^{\mu\nu\sigma}_{a}\right)\frac{\partial}{\partial p}
\nn\\
\hspace{-8ex}
&  &
-{f_{ab}} ^{c}\theta^{b}\left(p^{\mu\nu}_{c}\frac{\partial}{\partial p_{a}^{\mu\nu}}+\frac{1}{2!}\,p^{\mu\nu\sigma}_{c}\frac{\partial}{\partial p_{a}^{\mu\nu\sigma}} \right)
\,.
\eeq
By a straightforward calculation, it is possible to see that the vector field \eqref{GGSMPS} satisfies the identity
\begin{equation}\label{CCTGS}
\mathfrak{L}_{\xi_{\theta}^{Z}}\Theta_{{\mathrm{BF}}}=0\, ,
\end{equation}
which implies that the gauge symmetry group of the theory generates special covariant canonical transformations on $Z$, as expected. In a similar fashion, we also have that the vector field $\xi_{\chi}^{Z}\in\mathfrak{X}\left(Z\right)$ on $Z$ locally defined by
\beq
\label{GTSMPS}
\xi_{\chi}^{Z}
& := &
\left(D_{[\nu}\chi_{\sigma]}^{a}\frac{\partial}{\partial b_{\nu\sigma}^{a}}-D_{\nu}\left(\partial_{\mu}\chi^{a}_{\sigma}\right)p^{\mu[\nu\sigma]}_{a}\frac{\partial}{\partial p}-{f_{ab}}^{c}\chi^{b}_{\sigma}p^{\mu[\nu\sigma]}_{c}\frac{\partial}{\partial p^{\mu\nu}_{a}} \right)
\nn\\
& & +\epsilon^{\mu\nu\sigma\rho}\left(D_{\sigma}\chi_{a\rho}\frac{\partial}{\partial p^{\mu\nu}_{a}} +\frac{1}{2}f_{abc}\partial_{\sigma}\chi^{a}_{\rho} a_{\mu}^{b}a_{\nu}^{c}\frac{\partial}{\partial p}\right)\, ,
\eeq 
represents the vector field associated to the action of the topological symmetry of the BF theory on $Z$, which by means of the differential map $(\pi_{YZ})_\ast:TZ\rightarrow TY$ 
projects onto the generator of the topological symmetry on $Y$, $\xi_{\chi}^{Y}=
(\pi_{YZ})_\ast\left(\xi_{\chi}^{Z}\right)\in\mathfrak{X}\left(Y\right)$, and thus satisfies the condition 
\begin{equation}\label{CCTTS}
\mathfrak{L}_{\xi_{\chi}^{Z}}\Theta_{\mathrm{BF}}=d\alpha\, ,
\end{equation}
where $\alpha$ corresponds to the $3$-form on 
$Z$ given in \eqref{DFAlpha}. Note that the first line in \eqref{GTSMPS} represents the canonical lift of \eqref{IGGTS} from $TY$ to 
$TZ$ obtained by means of relation \eqref{CLMMPS}, while the second line corresponds to the part of the vector field $\xi_\chi^Z$ 
that brings the differential of $\alpha$ into play in equation~\eqref{CCTTS}. In consequence, the vector field \eqref{GTSMPS} corresponds to the $\alpha$-lift of $\xi_{\chi}^{Y}$ meaning that the vector field associated to \eqref{IGTS}  satisfies  condition \eqref{CCTTS} for the $3$-form $\alpha$ given in \eqref{DFAlpha}. This particular type of lifts of vector fields from $TY$ to $TZ$ are described and discussed in detail in \cite{DeLeon1}.

Finally, we are in the position to compute the covariant momentum maps associated to the gauge and topological symmetries of the BF theory. To this end, we start by taking into account the definition \eqref{CMM} which, together with the relations \eqref{CCTGS} and \eqref{CCTTS}, allows us to write the differential $3$-forms on $Z$ given by  
\beq
\label{CMMBFT}
\hspace{-5ex}
\mathcal{J}^{(\mathcal{M})}(\xi_{\theta})
&:= & 
\xi_{\theta}^{Z}\lrcorner\Theta_{\mathrm{BF}}
\nn\\
\hspace{-5ex}
& = & 
D_{\nu}\theta^{a}p^{\mu\nu}_{a}d^{3}x_{\mu}+\frac{1}{2!}\,{f^{a}}_{bc}b^{b}_{\nu\sigma}\theta^{c}p^{\mu\nu\sigma}_{a}d^{3}x_{\mu}\, ,
\nn\\
\hspace{-5ex}
\mathcal{J}^{(\mathcal{M})}(\xi_{\chi})
& := & 
\xi_{\chi}^{Z}\lrcorner\Theta_{\mathrm{BF}}-\alpha  
\nn\\
\hspace{-5ex}
& = & 
D_{[\nu}\chi^{a}_{\sigma]}p^{\mu\nu\sigma}_{a}d^{3}x_{\mu}-\frac{1}{2}\epsilon^{\mu\nu\sigma\rho}\chi_{a\nu}\left(da^{a}_{\rho}\wedge d^{2}x_{\mu\sigma}+{f^{a}}_{bc}a^{b}_{\sigma} a^{c}_{\rho}d^{3}x_{\mu}\right)\, ,
\eeq
which correspond to the covariant momentum maps of the BF theory. Thus, we have that by pulling-back the covariant momentum maps \eqref{CMMBFT} with the covariant Legendre transformation \eqref{CLT}, it is possible to recover the Noether currents \eqref{NCBFT} obtained in the geometric Lagrangian analysis of the BF theory. 
In particular, these covariant momentum maps 
will play an important role in order to 
establish a relation between the multisymplectic and the instantaneous Dirac-Hamiltonian formulations for the topological BF field theory.  We will discuss this in the following subsection.  

\subsection{Space plus time decomposition for non-Abelian BF theory}

In this subsection we will perform the space plus time decomposition for the non-Abelian topological BF theory. Our main aim is to recover the instantaneous Dirac-Hamiltonian analysis of the theory as developed, for example, in~\cite{Escalante1}, having as a 
starting point the multisymplectic formulation. To this end, we will follow the procedure described in subsection \eqref{STDCFT} above. To start, let $\Sigma_{t}$ be a $3$-dimensional Cauchy surface characterized by the level sets of the temporal coordinate $x^{0}$. Then, we introduce $\zeta_{X}:=\partial_{0}\in\mathfrak{X}\left(X\right)$ to be the generator of the slicing on the space-time manifold $X$. 
This in turn induces a generator of the slicing on the covariant configuration space of the BF theory, $Y$, defined through the relation
\begin{equation}\label{GIS}
\zeta_{Y}:=\zeta_{X}+\xi_{\theta}^{Y}+\xi_{\chi}^{Y},
\end{equation} 
which will be identified as the temporal direction of the covariant topological BF theory.  Here $\zeta_{X}$ must be thought of as a vector field on $Y$, while 
$\xi_{\theta}^{Y}\in\mathfrak{X}\left(Y\right)$ and $\xi_{\chi}^{Y}\in\mathfrak{X}\left(Y\right)$ correspond to the generators of the gauge and topological symmetries of the model, respectively. By construction, the vector field $\zeta_{Y}\in\mathfrak{X}\left(Y\right)$ leaves the action principle \eqref{BFT} invariant since its associated canonical prolongation from $TY$ to $T\left(J^{1}Y\right)$ corresponds to a Noether symmetry of the theory~\eqref{eq:noethersym}. Further, according to \eqref{TEFV}, the time evolution of the field variables is given by 
\beq
\label{TEFVBFT}
\dot{A}^{a}_{\mu}
&:= & 
\mathfrak{L}_{\zeta_{Y}}A^{a}_{\mu}=\partial_{0}A^{a}_{\mu}-D_{\mu}\theta^{a}
\,,\nn\\
\dot{B}^{a}_{\mu\nu}
&:= & 
\mathfrak{L}_{\zeta_{Y}}B^{a}_{\mu\nu}=\partial_{0}B^{a}_{\mu\nu}-{f^{a}}_{bc}B^{b}_{\mu\nu}\theta^{c}-2D_{[\mu}\chi^{a}_{\nu]}\, ,
\eeq
which corresponds to the time evolution of the field variables obtained by means of the extended Hamiltonian of the BF theory 
as discussed in~\cite{Escalante1}. 

Now we are in the position to introduce the space plus time decomposition of the BF theory at the Lagrangian level. To this end, given the compatible  
slicing $\left(\zeta_{X}\, \zeta_{Y}\right)$ of the covariant configuration space $\fibbun{Y}{X}$, we introduce $Y_{t}$ to denote the restriction of the manifold $Y$ to the Cauchy surface $\Sigma_{t}$. Then, the restricted jet bundle $J^{1}Y_{t}$ over $Y_{t}$ decomposes as $\beta_{\zeta_{Y}}:=J^{1}Y_{t}\rightarrow J^{1}(Y_{t})\times V\left(Y_{t}\right)$, which can be locally written as
\begin{equation}
\beta_{\zeta_{Y}}(x^{i},a^{a}_{\nu},b^{a}_{\nu\sigma},a^{a}_{\mu\nu},b^{a}_{\mu\nu\sigma})=(x^{i}, a^{a}_{\nu}, b^{a}_{\nu\sigma}, a^{a}_{i\nu}, b^{a}_{i\nu\sigma}, \dot{a}^{a}_{\nu}, \dot{b}^{a}_{\nu\sigma})\, .
\end{equation}
Hereinafter, we will use Latin letters $i,j$ and $k$ to denote spatial indices.

Thus, the instantaneous Lagrangian density of the theory, $
\mathcal{L}_{t,\zeta_{Y}}:J^{1}(Y_{t})\times V\left(Y_{t}\right)\rightarrow\Lambda^{3}\Sigma_{t}$, is defined by $\mathcal{L}_{t,\zeta_{Y}}:=\left(j^{1}\phi\circ i_{t}\right)^{*}\left(\zeta_{X}\lrcorner\,\mathcal{L}_{{\mathrm{BF}}}(\gamma)\right)$, with $j^{1}\phi\circ i_{\tau}$ denoting the restriction of the jet prolongation of the section $\phi\in\mathscr{Y}$ to the Cauchy surface $\Sigma_{t}$. In our case, the instantaneous Lagrangian of the non-Abelian BF theory, using the relations \eqref{TEFVBFT} and the definition \eqref{ILCFT}, can be explicitly written as
\begin{equation}
\label{eq:InstLag}
L_{t,\zeta_{Y}}\left(A^{a},B^{a}\right)=\frac{1}{2}\int_{\Sigma_{t}}\epsilon^{ijk}\left(B^{a}_{0i}F_{ajk}+\left(\dot{A}^{a}_{i}+D_{i}\xi^{a}-D_{i}A^{a}_{0}\right)B_{ajk}\right)d^{3}x_{0}\, ,
\end{equation}
where we have introduced the $3$-dimensional Levi-Civita alternating symbol $\epsilon^{ijk}:=\epsilon^{0ijk}$.
In particular, the instantaneous momenta associated to the fields $A^a$ and $B^a$, respectively, which corresponds to the image of the instantaneous Legendre transformation read as
\beq
\label{ILT}
\pi^{\nu}_{a}
& := & 
\frac{\partial L_{{\mathrm{BF}}}}{\partial \dot{A}^{a}_{\nu}} 
(A, B,\dot{A},\dot{B})
= 
\frac{1}{2}\delta^\nu_i \epsilon^{ijk} B_{ajk}   \,,
\nn\\
P^{\nu\sigma}_{a}
& := & 
\frac{\partial L_{{\mathrm{BF}}}}{\partial \dot{B}^{a}_{\nu\sigma}} 
(A, B,\dot{A},\dot{B})
=
0   \,.
\eeq
Here $L_{{\mathrm{BF}}}$ denotes the Lagrangian function associated to the system and corresponds to the integrand in~\eqref{eq:InstLag}.  These instantaneous 
momenta give rise to the primary constraint surface of the BF theory, that is, the surface $\mathscr{P}_{t}\subset T^{*}\mathscr{Y}_{t}$ defined by 
\begin{equation}\label{PCSet}
\mathscr{P}_{t}:=\{(A^{a},B^{a},\pi_{a},P_{a})\in T^{*}\mathscr{Y}_{t}\, |\, \pi^{0}_{a}=0\, , \pi^{i}_{a}-\frac{1}{2}\epsilon^{ijk}B_{ajk}=0\, , P^{0i}_{a}=0\, , P^{ij}_{a}=0\}\, ,
\end{equation}
where the instantaneous momenta $P^{\nu\sigma}_{a}$ associated to the field variables $B^{a}_{\nu\sigma}$ satisfy the condition $P^{\nu\sigma}_{a}=P^{[\nu\sigma]}_{a}$. 
Besides, the instantaneous momenta~\eqref{ILT} are not invertible in terms of the temporal derivatives of the fields, implying that the non-Abelian topological BF theory is described by a singular Lagrangian system according to \cite{QGS}.

Once we have introduced the space plus time decomposition at the Lagrangian level, we will proceed to perform this decomposition for the multimomenta phase-space of the BF theory. In particular, we are interested on 
studying the projected momentum maps associated to the generator of the gauge and topological symmetries of the theory and the canonical lift of $\zeta_{Y}$ to $Z$, which will coincide, as we will see below, with the generator of infinitesimal gauge transformations and the extended Hamiltonian of the BF theory obtained by means of the Dirac's algorithm in \cite{Escalante1}, respectively. To this end, we start by introducing the generator of the slicing on $Z$, that is, the lift of $\zeta_{Y}$
\eqref{GIS} from $TY$ to $TZ$, which explicitly reads 
\begin{equation}\label{GISMMPS}
\zeta_{Z}:=\zeta_{X}+\xi_{\theta}^{Z}+\xi_{\chi}^{Z}\, ,
\end{equation}
where we are considering $\zeta_{
X}$ as a vector field on $Z$, while the vector fields $\xi_{\theta}^{Z}$ and $\xi_{\chi}^{Z}$ correspond to the generators of the covariant canonical transformations associated to the action of the gauge and topological symmetries of the BF theory locally expressed in \eqref{GGSMPS} and \eqref{GTSMPS}, respectively. In particular, we introduce $\bar{\zeta}_{Z}:=\xi_{\theta}^{Z}+\xi_{\chi}^{Z}\in\mathfrak{X}\left(Z\right)$ to denote the part of the vector field $\zeta_Z$ that does not project, by means of the differential map $(\pi_{XZ})_\ast:TZ\rightarrow TX$, to a vector field on the manifold $X$ transverse to $\Sigma_{t}$.  
The vector field $\bar{\zeta}_{Z}$ will play an important role in what follows.

By definition~\eqref{GISMMPS}, we have that $\zeta_{Z}$ satisfies the condition $\mathfrak{L}_{\zeta_{Z}}\Theta_{\mathrm{BF}}=d\alpha$,  being $\alpha$ the $3$-form defined in \eqref{DFAlpha} and here considered as a differential form on $Z$, thus implying that $\zeta_{Z}$ generates covariant canonical transformations on $Z$.  Analogously, the vector field $\bar{\zeta}_{Z}$ also generates covariant canonical transformation on $Z$  since this last just corresponds to a superposition of the vector fields $\xi_{\theta}^{Z}$ and $\xi_{\chi}^{Z}$. Thus, $\zeta_{Z}$ and $\bar{\zeta}_{Z}$ have an associated covariant momentum map which projects by means of \eqref{PMM} to functions on $\mathscr{P}_{t}\subset T^{*}\mathscr{Y}_{t}$ and $T^{*}\mathscr{Y}_{t}$, respectively. 

As in subsection~\ref{STDCFT}, let $Z_{t}$ be the restriction of the fibre-bundle $\fibbun{Z}{X}:Z\rightarrow X$ to the Cauchy surface $\Sigma_{t}$, whose set of all smooth sections will be denoted by $\mathscr{Z}_{t}$. Thus, given $\sigma\in\mathscr{Z}_{t}$, the projected momentum map on the 
cotanget space $T^{*}\mathscr{Y}_{t}$ associated to the vector field $\bar{\zeta}_{Z}$  may be  defined, following~\eqref{PMM}, as
\begin{equation}\label{PMMGGS}
\begin{aligned}
\left\langle\mathscr{J}_{t}\left(A^{a}, B^{a}, \pi_{a}, P_{a}\right), \bar{\zeta}_{Z} \right\rangle&:=\int_{\Sigma_{t}}\sigma^{*}\left(\bar{\zeta}_{Z}\,\lrcorner\,\Theta_{\mathrm{BF}}-\alpha\right)\, ,\\
\end{aligned}
\end{equation}
which after identifying $\pi^{\nu}_{a}=p^{0\nu}_{a}\circ\sigma$ and $P^{\nu\sigma}_{a}=p^{0\nu\sigma}_{a}\circ \sigma$ can be explicitly written as
\beq
\hspace{-13ex}
\left\langle\mathscr{J}_{t}\left(A^{a}, B^{a}, \pi_{a}, P_{a}\right), \bar{\zeta}_{Z} \right\rangle
& = &
\int_{\Sigma_{t}}d^{3}x_{0}
\left\lbrace D_{0}\theta^{a}\pi^{0}_{a}
-\theta^{a}\left(D_{i}\pi^{i}_{a}+\frac{1}{2!}{f_{ab}}^{c}B^{b}_{ij}P^{ij}_{c}\right)
\right.
\, \nn\\
\hspace{-13ex}
& & 
\left.
+\left(2D_{[0}\chi^{a}_{i]}+{f^{a}}_{bc}B^{b}_{0i}\theta^{c}\right)P^{0i}_{a}
-\chi^{a}_{i}\left(\frac{1}{2}\epsilon^{ijk}F_{ajk}-D_{j}P^{ij}_{a}\right) 
\right\rbrace \, ,
\nn
\eeq
where we have performed some integration by parts and avoid terms on the boundary of the Cauchy surface. Then, by introducing the parameters 
\beq
\epsilon^{a}_{0}
&:=& 
D_{0}\theta^{a}\, ,
\nn\\
\epsilon^{a}_{i}
&:=& 
2D_{[0}\chi^{a}_{i]}+{f^{a}}_{bc}B^{b}_{0i}\theta^{c}\, ,
\nn\\
\lambda^{a}
&:=& 
-\theta^{a}\, ,
\nn\\
\lambda^{a}_{i}
&:=& 
-\chi^{a}_{i}\, ,
\eeq
the projected momentum map associated to $\bar{\zeta}_{Z}$ can be expressed in the following fashion
\beq
\hspace{-7ex}
\left\langle\mathscr{J}_{t}\left(A^{a}, B^{a}, \pi_{a}, P_{a}\right), \bar{\zeta}_{Z} \right\rangle 
& = & 
\int_{\Sigma_{t}}\!d^{3}x_{0} \left\lbrace \epsilon^{a}_{0}\pi^{0}_{a}
+
\epsilon^{a}_{i}P^{0i}_{a} 
+
\lambda^{a}\!\left(D_{i}\pi^{i}_{a}+\frac{1}{2!}{f_{ab}}^{c}B^{b}_{ij}P^{ij}_{c}\right)
\right.
\nn\\
\hspace{-7ex}
& & 
\left. 
+
\lambda^{a}_{i}\!\left(\frac{1}{2}\epsilon^{ijk}F_{ajk}-D_{j}P^{ij}_{a}\right) \right\rbrace \, ,
\nn
\eeq
which corresponds precisely to the generator of the infinitesimal gauge transformations in Dirac's terminology of the non-Abelian topological BF theory obtained by means of the Dirac's algorithm in \cite{Escalante1}.  
Further, as described above in subsection~\ref{STDCFT}, the vanishing of the projected momentum map \eqref{PMMGGS} is related to the first class constrains, within Dirac's terminology \cite{QGS}, of the BF theory in the instantaneous Dirac-Hamiltonian approach. To see this, we start by computing the zero level
set of the projected momentum map \eqref{PMMGGS}, that is, the surface on $T^{*}\mathscr{Y}_{t}$ defined by ${\mathscr{J}_{\mathrm{BF}}}_{t}^{-1}\left(0\right):=\{\left(A^{a},B^{a},\pi_{a},P_{a}\right)\subset T^{*}\mathscr{Y}_{t}\left|\right. \left\langle\mathscr{J}_{t}\left(A^{a}, B^{a}, \pi_{a}, P_{a}\right), \bar{\zeta}_{Z} \right\rangle=0\}$, which can be explicitly written, since the parameters $\epsilon^{a}_{0}, \epsilon^{a}_{i}, \lambda^{a}$ and $\lambda^{a}_{i}$ are arbitrary functions on $\Sigma_{t}$, as
\beq
\label{FCCBFT}
\hspace{-11ex}
{\mathscr{J}_{\mathrm{BF}}}_{t}^{-1}\left(0\right)
& = & 
\left\lbrace\left(A^{a},B^{a},\pi_{a},P_{a}\right)\subset T^{*}\mathscr{Y}_{t}\left|\right. \pi^{0}_{a}=0\, ,~ P^{0i}_{a}=0\, ,
D_{i}\pi^{i}_{a}+\frac{1}{2!}{f_{ab}}^{c}B^{b}_{ij}P^{ij}_{c}=0\, ,
\right.
\nn\\
\hspace{-11ex}
& & 
\left.
\frac{1}{2}\epsilon^{ijk}F_{ajk}-D_{j}P^{ij}_{a}=0\right\rbrace\, ,
\eeq
giving rise to the constrained surface on the instantaneous phase-space $T^{*}\mathscr{Y}_{t}$ defined by the first class constraints of the BF theory as obtained in \cite{Escalante1}.

Besides, since $\zeta_{Z}$ projects into the vector field transverse to $\Sigma_{t}$, $\zeta_X$, by means of the differential map $(\fibbun{Z}{X})_\ast$, the covariant momentum map associated to the vector field $\zeta_{Z}$ projects into a well defined function on the image of the Legendre transformation, that is, on the primary constraint set $\mathscr{P}_{t}\subset T^{*}\mathscr{Y}_{t}$~\cite{GIMMSY2, DeLeon2}.  
This function will be related, by means of \eqref{IHT}, with the instantaneous extended Hamiltonian function of the BF theory. To see this, let $\sigma$ be the canonical lift associated to $\left(A^{a}, B^{a}, \pi_{a}, P_{a}\right)\in\mathscr{P}_{t}$, that is, $\sigma\in \mathscr{N}_{t}$. Then, we introduce the following function on $\mathscr{P}_{t}$
\begin{equation}
H_{t,\zeta_{Z}}\!\left(A^{a}, B^{a}, \pi_{a}, P_{a}\right):=-\int_{\Sigma_{t}}\sigma^{*}\left(\zeta_{Z}\,\lrcorner\,\Theta_{\mathrm{BF}}-\alpha\right)\, ,
\end{equation}
which after some straightforward calculations may be explicitly written as 
\beq
\hspace{-7ex}
H_{t, \zeta_{Z}}\!\left(A^{a}, B^{a}, \pi_{a}, P_{a}\right)
& = &
\int_{\Sigma_{\tau}}\!d^{3}x_{0} \left\lbrace -A^{a}_{0}D_{i}\pi^{i}_{a}-\frac{1}{2}\epsilon^{ijk}B_{a0i}F^{a}_{jk}+\frac{1}{2!}\partial_{0}B^{a}_{ij}P^{ij}_{a}
\right.
\nn\\
\hspace{-7ex}
& & 
+
\dot{A}^{a}_{0}\pi^{0}_{a}\!+\!\dot{B}^{a}_{0i}P^{0i}_{a}\!+\!\xi^{a}\!\left(D_{i}\pi^{i}_{a}+\frac{1}{2!}{f_{ab}}^{c}B^{b}_{ij}P^{ij}_{c}\right)
\nn\\
\hspace{-7ex}
& &
\left. 
+ \chi^{a}_{i}\!\left(\frac{1}{2}\epsilon^{ijk}F_{ajk}-D_{j}P^{ij}_{a}\right)\right\rbrace\, .
\eeq
In particular, by using the field equations for the BF theory~\eqref{BFFE}, it is possible to obtain the following relations
\begin{equation}
\label{eq:LagMultipliers}
\partial_{0}B^{a}_{ij}=2D_{[i}B^{a}_{0|j]}-{f^{a}}_{bc}A^{b}_{0}B^{c}_{ij}\, ,
\end{equation}
that allow us to explicitly write the instantaneous Hamiltonian function $H_{t, \zeta_{Z}}$ on $\mathscr{P}_{t}$ 
in correspondence to the extended Hamiltonian of the non-Abelian topological BF theory obtained by means of the Dirac's algorithm for constrained systems as described in~\cite{Escalante1}.  
In order to establish this correspondence we note that the factor $2D_{[i}B^{a}_{0|j]}-{f^{a}}_{bc}A^{b}_{0}B^{c}_{ij}$ introduced in~\eqref{eq:LagMultipliers} may be identified with the fixed Lagrange multiplier that enforces the primary constraint $P^{ij}_{a}\approx 0$ into the extended Hamiltonian of the theory.  This Lagrange multiplier is fixed, within Dirac's formalism, by imposing the so-called consistence conditions on the primary constraint set of the BF theory, described above as~\eqref{PCSet}. 

In particular, by taking into account that the surface on $T^{*}\mathscr{Y}_{t}$ given by \eqref{FCCBFT} corresponds to the surface defined by the first class constraints of the BF theory, we have that the primary constraints \eqref{PCSet} of the system, which does not appear in the set \eqref{FCCBFT} must be second class constraints within Dirac's terminology, that is, the functions $\pi^{i}_{a}-\frac{1}{2}\epsilon^{ijk}B_{ajk}=0$ and $P^{ij}_{a}=0$ on $T^{*}\mathscr{Y}_{t}$ correspond to the second class constraints of the non-Abelian topological BF theory. Also, the first class constraints \eqref{FCCBFT} are not all independent since the field strength $F^{a}_{\mu\nu}$ satisfies the Bianchi identity $\epsilon^{\mu\nu\sigma\rho}D_{\nu}F^{a}_{\sigma\rho}=0$, which reduces the number of linearly independent first class constraints $\frac{1}{2}\epsilon^{ijk}F_{ajk}-D_{j}P^{ij}_{a}=0$ to $2N$. Bearing this in mind, it is possible to see that, according to \cite{QGS}, the number of local degrees of freedom per point in the instantaneous phas-space of the BF theory corresponds to $N\times\left( 2\cdot10 -2\cdot7-6 \right)=0$, which implies that, the non-Abelian BF theory defined by \eqref{BFT} corresponds to a topological field theory \cite{Escalante1, Montesinos1}.

To summarize, by performing the space plus time decomposition of the multisymplectic formulation of the topological BF model we have successfully recovered not only the first and second class constraints, 
but also the generator of the infinitesimal gauge transformations together with the instantaneous extended Hamiltonian of the theory.   Our 
results are completely consistent with those obtained within the instantaneous Dirac-Hamiltonian analysis of the model.

\section{Conclusions}
\label{sec:Conclu}

In this paper we have analyzed the $4$-dimensional non-Abelian topological BF field theory within the geometric and covariant Lagrangian and multisymplectic formalisms for classical field theory. We have focused our study to the analysis of the gauge and the topological symmetries of the model
BF model.
On the one hand, at the Lagrangian level we found that the gauge symmetry corresponds to a natural symmetry of the system while the topological symmetry represents a Noether symmetry for this topological field theory. Following the geometric Lagrangian approach, we obtained the field equations and the Noether currents associated to the non-Abelian BF theory, obtaining results that are consistent with those reported in~\cite{Montesinos1}. In particular, we found that the Noether  current associated to the gauge symmetry group of the theory integrated over a Cauchy surface vanishes by requiring that gauge arbitrary functions on space-time to be of compact support on the boundary of the Cauchy surface, while, in an analogous manner,  the Noether current associated to the topological symmetry is trivially zero on each point of the space-time manifold. Indeed, this is consistent with the behavior for a generic 
field theory with gauge symmetries~\cite{Sardanashvily1, Avery, Lee}.
On the other hand, within the multisymplectic approach, we found that the gauge symmetry group of the BF theory acts on the associated multimomenta phase-space by special covariant canonical transformations, while the topological symmetry of the model acts on the multimomenta phase-space by covariant canonical transformations.  We also describe the existence of covariant momentum maps associated to the actions of the gauge and the topological symmetries of the non-Abelian BF theory. This 
covariant momentum maps for the BF theory allowed us to recover, for any solution of the BF field equations, the Noether currents associated to this topological field theory.  

In addition, we performed the space plus time decomposition for the non-Abelian topological BF field theory at the Lagrangian and the multisymplectic levels. To this end, working in adapted coordinates, we foliated the space-time manifold into Cauchy surfaces and 
introduced the generators of the associated slicings on both, the space-time manifold and the covariant configuration space of the theory, respectively. Once the temporal direction has been fixed by this slicing and after appropriately decomposing the jet bundle of the theory, we computed the instantaneous Lagrangian of the non-Abelian BF theory on the space of Cauchy data, which allowed us to obtain, by means of the instantaneous Legendre transformation, the instantaneous momenta variables of the model, which gave rise to the primary constraint set of the BF theory.
Besides, within the multisymplectic approach, we also constructed the generator of the slicing on the multimomenta phase-space of the theory. In particular, we showed at the multiisymplectic level that the superposition of the covariant momentum maps associated to the gauge and topological symmetries of the BF theory projects into the instantaneous phase-space generator of
infinitesimal gauge transformations of the system and, and we also realized that 
the zero level set of the above projected momentum map coincides with the surface on the instantaneous phase-space defined by the first class constraints of the non-Abelian topological BF field theory, which arise within the Dirac-Hamiltoniana analysis of the model as described in~\cite{Escalante1, DeGracia}. Furthermore, we found that since the generator of the slicing on the multimomenta phase-space generates covariant canonical transformations, its associated covariant momentum map projects to the extended Hamiltonian of this topological field theory also obtained by means of the Dirac's algorithm in the instantaneous canonical analysis~\cite{Escalante1}.

With all these results, we have showed that the multisymplectic approach allows to describe in a covariant, consistent and elegant way the features of the $4$-dimensional non-Abelian topological BF field theory. In particular, we have found that it is possible to recover the 
instantaneous Dirac-Hamiltonian formulation of the theory by performing the space plus time decomposition fo the multisymplectic phase-space
as developed in \cite{GIMMSY1, GIMMSY2, DeLeon1, DeLeon2, Fischer, Gotay1}.  
To our knowledge, this is the first non-trivial physical model associated to General Relativity for which both natural and Noether symmetries have been analyzed at the multisymplectic level.  The vast majority of the 
examples explored so far only considered natural symmetries avoiding the 
construction of the so-called $\alpha$-lifts associated to 
Noether symmetries.  Those $\alpha$-lifts, as we showed for this example and as discussed extensively in~\cite{DeLeon1}, 
play an important role in order to construct the correct covariant momentum 
maps associated to Noether symmetries.   From the physical point of view, the understanding of the effect of these $\alpha$-lifts is completely relevant to recover the correct generators of gauge transformations for 
a given field theory.   Our intention is to explore in detail, for some physical gauge models  associated to General Relativity, the applicability 
of the covariant formalism here described and, in particular, the way to 
recover the generators of the gauge transformations in comparison to the 
instantaneous Dirac-Hamiltonian approach.  This will be done elsewhere.

\section*{Acknowledgments}
The authors would like to acknowledge financial support from CONACYT-Mexico
under project CB-2014-243433.  A.R.-L. was also supported by a CONACYT-Mexico Postgraduate Fellowship.


\section*{References}





\begin{thebibliography}{99}


\bibitem{Cattaneo} A. S. Cattaneo, P. Cotta-Ramusino, J. Froehlich and M. Martellini, \textit{Topological BF Theories in 3 and 4 Dimensions}, J. Math. Phys. \textbf{36}, 6137-6160 (1995), \texttt{arXiv:hep-th/9505027v2}.

\bibitem{Horowitz} G. Horowitz, \textit{Exactly Soluble Diffeomorphism Invariant Theories}, Commun. Math. Phys. \textbf{125}, 417–437 (1989).

\bibitem{Birmingham} D. Birmingham, M. Blau, M. Rakowski and G. Thompson, \textit{Topological Field Theory}, Phys. Rep. \textbf{209}, 129-340 (1991).

\bibitem{Plebanski} J. F. Plebanski, \textit{On the separation of Einsteinian substructures}, J. Math. Phys. \textbf{18}, 2511 (1977).

\bibitem{Speziale} L. Freidel and S. Speziale, \textit{On the Relations between Gravity and BF Theories}, SIGMA \textbf{8}, 032 (2012), \texttt{arXiv:1201.4247[gr-qc]}.

\bibitem{Montesinos2} M. Celada, D. González and M. Montesinos, \textit{BF gravity}, Class. Quantum Grav. \textbf{33}, 213001 (2016), \texttt{arXiv:1610.02020[gr-qc]}.

\bibitem{Sardanashvily1} G. Giachetta, L. Mangiarotti and G. Sardanashvily, \textit{Advanced Classical Field Theory}, (World Scientific Publishing Co. Pte. Ltd., 2009).

\bibitem{Spinfoams1} J. Engle, R. Pereira and C. Rovelli, \textit{Flipped spinfoam vertex and loop gravity}, Nucl. Phys. \textbf{B} 798, 251--290 (2008), \texttt{arXiv:0708.1236 [gr-qc]}.

\bibitem{Spinfoams2} J. C. Baez, \textit{An introduction to spin foam models of quantum gravity and BF theory}, Lect. Notes Phys. \textbf{543}, 25–94 (2000), \texttt{arXiv:gr-qc/9905087}.

\bibitem{DeGracia} G. B. de Gracia, B. M. Pimentel and C. E. Valcárcel, \textit{Hamilton--Jacobi analysis of the four dimensional BF model with cosmological term}, Eur. Phys. J. Plus \textbf{132}, 438 (2017), \texttt{arXiv:1702.00863v1 [hep-th]}. 

\bibitem{Montesinos3} M. Mondragón and M. Montesinos, \textit{Covariant canonical formalism for four-dimensional BF theory}, J. Math. Phys. \textbf{47}, 022301 (2006), \texttt{arXiv:gr-qc/0402041}.

\bibitem{Angel}
J.~Berra-Montiel, 
A.~Molgado and 
A.~Rodr\'iguez-L\'opez,
\emph{Polysymplectic formulation for BF gravity with Immirzi parameter}, 
Class.~Quantum Grav. 
{\bf 36} 
115003 
(2019),
\texttt{arXiv:1901.11532v2 [gr-qc]}.

\bibitem{Montesinos1} M. Montesinos, \textit{Noether currents for BF gravity}, Class. Quantum Grav. \textbf{20}, 3569–3575 (2003). 

\bibitem{Escalante1}  
A. Escalante and I. Rubalcava-García, \textit{A pure Dirac's canonical analysis for four-dimensional BF theories}, Int. J. Geom. Methods Mod. Phys. \textbf{09}, 1250053 (2012), \texttt{arXiv:1107.4421v1}.



\bibitem{Forger1} M. Forger and S. V. Romero, \textit{Covariant Poisson Brackets in Geometric Field Theory}, Commun. Math. Phys. \textbf{256}, 375-410 (2005), \texttt{arXiv:math-ph/0408008v1}.

\bibitem{GIMMSY1} M. J. Gotay, J. Isenberg, J. Marsden and R. Montgomery, \textit{Momentum maps and classical
relativistic fields. Part I: Covariant field theory}, (1998), 
\texttt{arXiv:physics/9801019v2 [math-ph]}.

\bibitem{Crampin} J.~F.~Cariñena, M.~Crampin and L.~A.~Ibort, \textit{On the multisymplectic formalism for first
order field theories}, Diff. Geom. App. \textbf{1}, 345--374 (1991). 

\bibitem{DeLeon1} M. De León, D. M. De Diego and A. Santamaría--Merino, \textit{Symmetries in Classical Field Theory}, Int. J. Geom. Meth. Mod. Phys. \textbf{01}, 651-710 (2004), \texttt{arXiv:math-ph/0404013v2}.

\bibitem{Gotay1} M. Gotay, \textit{A multisymplectic framework for classical field theory and the calculus of variations II: space + time decomposition}, Diff. Geom. Appl. \textbf{1}, 375–390 (1991).

\bibitem{Forger2} M. Forger and M. O. Salles, \textit{On Covariant Poisson Brackets in Classical Field Theory}, J. Math. Phys. \textbf{56}, 102901 (2015), \texttt{ arXiv:1501.03780v1}.

\bibitem{GIMMSY2} M. J. Gotay, J. Isenberg and J. Marsden, \textit{Momentum maps and classical
relativistic fields. Part II: Canonical Analysis of Field Theories}, (2004), \texttt{arXiv:math-ph/0411032}.

\bibitem{Fischer} E. Binz, J. Sniatycki and H. Fischer, \textit{Geometry of classical fields}, North-Holland Mathematics Studies \textbf{154}, (North-Holland Publishing, 1988). 

\bibitem{QGS} M. Henneaux and C. Teitelboim, \textit{Quantization of gauge systems} (Princeton University Press, 1994).

\bibitem{Avery} S. G. Avery and B. U. W. Schwab, \textit{Noether's second theorem and Ward identities for gauge symmetries}, J. High Energ. Phys. \textbf{2016}, 31 (2016), \texttt{arXiv:1510.07038 [hep-th]}.

\bibitem{Lee} J. Lee and R. M. Wald, \textit{Local symmetries and constraints}, J. Math. Phys. \textbf{31}, 725--743 (1990).

\bibitem{DeLeon2} C. M. Campos, M. De León, D. M. De Diego and M. Vaquero, \textit{Hamiltonian–Jacobi Theory in Cauchy Data Space}, Rep. Math. Phys. \textbf{76}, 359–387 (2015), \texttt{arXiv:1411.3959v1 [math-ph]}.










\bibitem{Saunders} D. J. Saunders, \textit{The Geometry of Jet Bundles} (Lecture Notes Series vol 142) (Cambridge University Press, 1989).


\bibitem{Vankerschaver} J. Vankerschaver, \textit{The momentum map for nonholonomic field theories with symmetry}, Int. J. Geom. Meth. Mod. Phys. \textbf{2}, 1029--1042 (2005), \texttt{arXiv:math-ph/0507059v1}.










\end{thebibliography}
\end{document}